\documentclass[universe,article,accept,pdftex,moreauthors]{Definitions/mdpi} 
\firstpage{1} 
\pubvolume{1}
\issuenum{1}
\articlenumber{0}
\pubyear{2025}
\copyrightyear{2025}
\externaleditor{Academic Editor: Diego Turrini}
\datereceived{31 October 2024} 
\daterevised{20 December 2024} 
\dateaccepted{23 December 2024} 
\datepublished{25 December 2024} 
\hreflink{https://doi.org/10.3390/\linebreak
universe11010001} 

\Title{Effects of Thermodynamics on the Concurrent Accretion and Migration of Gas Giants in Protoplanetary Disks}

\TitleCitation{Effects of Thermodynamics On the Concurrent Accretion and Migration of Gas Giants in Protoplanetary Disks}

\usepackage{bm}

\usepackage{amsmath}
\usepackage{graphicx}
\newcommand\aj{Astron. J.}
\newcommand\araa{Ann. Rev. Astron. Astrophys.}
\newcommand\apj{Astrophys. J.}
\newcommand\apjl{Astrophys. J. Lett.}     
\newcommand\apjs{Astrophys. J. Suppl. Ser.}
\newcommand\aap{Astron. Astrophys.}
\newcommand\icarus{Icarus}
\newcommand\mnras{Mon. Not. R. Astron. Soc.}
\newcommand\nat{Nature}

\newcommand\maps{M\&PS}

\renewcommand{\vec}[1]{{\bm #1}}

\Author{Hening Wu $^{1}$\orcidA{}, Ya-Ping Li $^{2}$\orcidB{}}

\AuthorNames{Hening Wu, Ya-Ping Wu}

\AuthorCitation{Wu, H.; Li, Y.-P.}

\address{%
$^{1}$ \quad Department of Physics and Astronomy, University of Nevada, Las Vegas, 4505 S. Maryland Parkway, Las Vegas, NV 89154, USA; hening.wu@unlv.edu\\
$^{2}$ \quad Shanghai Astronomical Observatory, Chinese Academy of Sciences, Shanghai 200030, People’s Republic of China; liyp@shao.ac.cn}

\corres{Correspondence: hening.wu@unlv.edu (H.W.); liyp@shao.ac.cn (Y.P.L.)}

\abstract{Accretion and migration usually proceeds concurrently for giant planet formation in the natal protoplanetary disks. Recent works indicate that the concurrent accretion onto a giant planet imposes significant impact on the planetary migration dynamics in the isothermal regime.
In this work, we carry out a series of 2D global hydrodynamical simulations with \texttt{Athena++} to explore the effect of thermodynamics on the concurrent accretion and migration process of the planets in a self-consistent manner. The thermodynamics effect is modeled with a thermal relaxation timescale using a $\beta$-cooling prescription.
Our results indicate that radiative cooling has a substantial effect on the accretion and migration processes of the planet.
As cooling timescales increase, we observe a slight decrease in the planetary accretion rate, and a transition from the outward migrating into inward migration. This transition occurs approximately when the cooling timescale is comparable to the local dynamical timescale ($\beta\sim1$), which is closely linked to the asymmetric structures from the circumplanetary disk (CPD) region. The asymmetric structures in the CPD region which appear with an efficient cooling provide a strong positive torque driving the planet migrate outward. However, such a positive torque is strongly suppressed, when the CPD structures tend to disappear with a relatively long cooling timescale ($\beta\gtrsim10$).
Our findings may also be relevant to the dynamical evolution of accreting stellar-mass objects embedded in disks around active galactic nuclei.
}


\begin{document}

\setcounter{section}{-1} 
\section{INTRODUCTION} \label{sec:intro}

With several exoplanet populations have been detected so far, the discovery of the close-in super-Earths \citep{Borucki2011} and hot Jupiters \citep{1995Natur.378..355M} has prompted the theoretical proposal that planets may migrate significantly in the protoplanetary disk (PPD) phase \citep{1986ApJ...309..846L, Levison2007}, although it has been suggested that the giant planet could also migrate due to a more populated asteroid belt \citep{Clement2020}. Another compelling evidence for disk-driven migration is the discovery of multiple planets in or near mean motion resonant configurations \citep{Lissauer2011, Goldreich2014,Xu2018,Yang2024}. 
The migration of giant planets has a substantial dynamic impact on the formation regions and final locations of minor bodies \citep[e.g.,][]{Brownlee2006,Huang2024}.
For the migration of a low-mass protoplanet, the gravitational interaction with the PPD is usually the dominant factor which leads to the classical type I migration \citep{1997Icar..126..261W}. The torque exerted onto the protoplanet can be divided into two parts, the wave torque \citep{1980ApJ...241..425G, 1993ApJ...419..155A}, and the corotation torque or horseshoe drag \citep{1991LPI....22.1463W,Masset2001}, which originate from the asymmetries in different regions of the PPD and usually drives the planet to migrate inward \citep{1984ApJ...285..818P, Tanaka2002, Kley2012,Paardekooper2023}. When the protoplanet's mass is high enough to modify the density distribution of the PPD, the tidal interaction between the protoplanet and the disk leads to the opening of a gap around the protoplanet's orbit, and the migration is in the realm of type II, which is believed to be locked to the viscous drift velocity of the background disk \citep{1980ApJ...241..425G,1986ApJ...307..395L}. 

In the meanwhile, under the widely adopted core accretion model for planet formation, the dynamical runaway gas accretion occurs at a "critical core mass" above which the envelope fails to maintain hydrostatic equilibrium
\citep{1986Icar...67..391B, 1996Icar..124...62P, 2004ApJ...604..388I,Chen2020}.
In the phase of the dynamical gas accretion, the gas supplied by the PPD, due to the conservation of angular momentum, could form a disk around the protoplanet, known as the circumplanetary disk (CPD) \citep{Machida2008, Tanigawa2012, Fung2019, Li2023}.
The accretion onto the protoplanets sensitively depend on the structures of the CPD, which finally determine the asymptotic mass of mature planetary system \citep[e.g.,][]{Kley2001,DAngelo2003,Machida2010,Bodenheimer2013,Choksi2023,Li2021a,Li2023}.
The presence of the deep gap and the strong tidal barrier effect in the type II regime would prevent the gas from reaching the protoplanets' vicinity \citep{1980ApJ...241..425G, 1986ApJ...307..395L}, thus impeding their growth \citep[e.g.,][]{Dobbs-Dixon2007,Li2021a}.
However, further studies have shown that the gas in the PPD could still pass through the gap through the horseshoe orbit, so that the protoplanet could accrete gas anyway \citep{Duffell2014,Fung2014,DurmannKley2015,Robert2018,Chen2020b,Li2023}. Therefore, a new paradigm for the type II migration emerges, which suggests that the type II migration rate is not at the viscous velocity of the disk, but shows a similar scaling law as the type I case after considering the steady gap structure \citep{Kanagawa2018}.

It has been shown that the accretion process in the CPD is mainly contributed by spiral shock dissipation, which is sensitive to the disk thermodynamics \citep{Zhu2016}. While most of the studies for the CPD structures and the gas dynamical accretion adopted an isothermal equation of state (EOS), there have been some works including non-isothermal effect and radiation hydrodynamics for the CPD formation \citep[e.g.,][]{Szulagyi2016,Fung2019,Krapp2024}. Different treatments of the thermodynamics could also influence the circumbinary disk accretion and dynamics \citep[e.g.,][]{Li2022b,Sudarshan2022,Wang2023}. 
Furthermore, the disk thermodynamics could alter the global disk structure \citep{Miranda2020,Zhang2020,Ziampras2023,Zhang2024} and thus the planet migration dynamics significantly \citep{Masset2009,Paardekooper2010,Lega2014}. However, the accretion and migration of planets in these works are typically studied in isolation. In this sense, the coupled evolution between the CPD accretion and planetary accretion are not treated in a self-consistent manner, as also highlighted in \citet{Li2024}.

Recently, it has been found that the concurrent accretion would imposes strong impact on the migration of the planets in PPDs when taking into account of a fully self-consistent treatment for an accreting planet \citep{Li2024,Laune2024}. Especially, \citet{Li2024} found that the structures of CPD play an important role on the migration of the planets in the isothermal limit. As the thermodynamics of the disk could modify the CPD structures significantly, it is thus still unclear how different thermodynamics affect the concurrent accretion and migration of planet in PPDs, which is the main goal of this work. Since there have been some theoretical arguments suggesting that there could a population of stellar-mass objects (e.g., stellar-mass black holes, stars) embedded in active galactic nucleus (AGN) disks,  our studies should also be relevant to the accretion and migration of stellar-mass objects embedded in AGN disks.

This rest part of the paper is organized as follows. We present the numerical methods in our hydrodynamical simulations in Section \ref{sec:num_method}. The results are shown in Section \ref{sec:sim_results}. The conclusions and discussions are presented in Section \ref{sec:conclusions}.

\section{NUMERICAL METHODS} \label{sec:num_method}

\subsection{Hydrodynamical Model} \label{subsec:hydro_model}

We simulate globally an accreting planet embedded in a thin and non-self-gravitating disk  using  \texttt{Athena++} \citep{2020ApJS..249....4S}. Most of our simulation setup are similar to  \citet{Li2024}. Here we only briefly describe the main physics of our models.
The simulations are performed in a cylindrical coordinate system ($r,\phi$), and 
the origin of the coordinate system is located at the position of the central star with 
mass $M_{*}$.
We initialize a constant aspect ratio of the PPD as $h \equiv \frac{H}{r} = 0.05$ over the whole disk,
where $H$ is the disk scale height. This leads to an initial  temperature profile of $T_{\rm disk}\propto r^{-1.0}$. The local sound speed is thus $c_{\rm s}=h v_{\rm K}$, where $v_{\rm K}$ is the local Keplerian velocity of the disk. The disk has an axis-symmetric initial profile of $\Sigma(r) = \Sigma_{0} (\frac{r}{r_{0}})^{p}$ with $p=-0.5$, where $r_{0}$ is the typical length scale of the disk, and $\Sigma_{0}$ is the surface density at $r_{0}$. Due to the high uncertainty from the observations, the exponents of the temperature and surface density profiles exhibit considerable variations for different PPDs 
at different radial locations \citep{Andrews2020,Miotello2023}. Therefore, we adopt specific exponents of the profiles to maintain a steady-state accretion disk, remaining within the acceptable range of observational uncertainties.

The planet is fixed at a circular orbit with a semi-major axis of $a_{\rm p}=r_{0}$ in a corotating frame with the planet so that the planet's azimuthal angle fixed at $\phi_{\rm p}=0$. We consider a planet with 1 Jupiter mass ($m_{\rm p}=M_{\rm J}$) which corresponds to a mass ratio between the planet and the solar-type star of $q\equiv\frac{m_{\rm p}}{M_{*}}=0.001$. The gravitational potential of the planet at $\vec{r}_{\rm p}$ is in the form,

\begin{equation}\label{equ:gra_poten}
\Phi_{\rm p} = -\frac{Gm_{\rm p}}{(|\vec{r_{\rm p}}-\vec{r}|^{2}+s^{2})^{1/2}} + q \Omega_{\rm p}^{2} \vec{r_{\rm p}} \cdot \vec{r},
\end{equation}
where $s$ indicates the softening length of the potential which is set as  $0.1\ r_{\rm h}$ unless otherwise stated, and $r_{\rm h}$ is the Hill radius of the planet. Other softening lengths have also been tested for the convergence, as shown in the Appendix \ref{subsec:con_test}.

We adopt a Shakura-Sunyaev viscosity prescription $\nu_{g}=\alpha c_{\rm s} H$ with a constant $\alpha=0.01$ across the whole disk \citep{1973A&A....24..337S}. 
A relative large viscosity parameter is to ensure the disk reach the quasi-viscous steady state within a feasible computation time. The disk, covering a radial range of $[r_{\rm in},r_{\rm out}]=[0.3, 3.5]\ r_{0}$, is resolved with a uniform base grid of 80 and 256 cells in radial and azimuthal directions, respectively.
We have implemented four levels of static mesh refinement within a distance of $1\ r_{\rm h}$ from the planet to effectively resolve the CPD region.
With such a grid resolution, the Hill radius can be resolved by about 20 cells in each dimension. A test with different grid resolutions have been performed as shown in the Appendix~\ref{subsec:con_test} to confirm the convergence of our simulation results.
For all of our simulations, we take the natural units of $G=M_{*}=r_{0}=1.0$, and the simulation results are scale-free in a non-self-gravitating disk.

\subsection{Cooling of the Disk} \label{subsec:cool_disk}
To explore the effects of thermodynamics, we conduct simulations with locally isothermal and adiabatic EOSs with $\beta-$cooling, as adopted in previous works \citep[e.g.,][]{Zhu2015,Zhang2020,Zhang2024}. The EOS for the isothermal case is

\begin{equation}
P = c_{\rm s}^{2}\Sigma. \label{EOS_iso}
\end{equation}
The adiabatic EOS is

\begin{equation}
P = (\gamma - 1)e, \label{EOS_adi}
\end{equation} 
where $P$ is the pressure, and $e$ is the specific internal energy of the gas. The sound speed in this case is defined as $c_{\rm s}=\sqrt{\gamma P/\Sigma}$. In our simulations, we take $\gamma=1.4$ for Equation~\ref{EOS_adi}. In addition, We perform the adiabatic EOS simulations with a simple thermal relaxation process, such that the gas internal energy is relaxed towards its initial state on a timescale controlled by the parameter $\beta$,

\begin{equation}
\frac{dE(t)}{dt} = -\frac{\Omega}{\beta}(E(t) - E(0)),
\end{equation}
where $\Omega$ is the local orbital frequency of the disk. In the calculation, this is equivalent to the following equation in which $E$ is transformed after each time step as follows,

\begin{equation}
E_{t+dt} = E_{t}(1 - \frac{\Omega}{\beta}dt) + E_{0}\frac{\Omega}{\beta}dt,
\end{equation}
where $E_{t+dt}$ and $E_{t}$ are the internal energy of the gas in the grid before and after each time step, $E_{0}$ indicates the initial state. So the local cooling timescale $t_{\rm cool}$ of the system is $\beta / \Omega$. The smaller the value of $\beta$, the closer the disk is to the locally isothermal condition. 
In contrast, a higher value of $\beta$ brings the disk closer to the adiabatic approximation. 
Assuming a Minimum Mass Solar Nebula, the cooling timescale is shorter (e.g., $\beta\simeq10^{-2}$ at 100 au) at the outer disk and much longer at the inner disk (e.g., $\beta\simeq10^{5}$ at 1 au) \citep{Zhu2015}.
In our numerical simulations, we examine $\beta$ values across a broad range from $10^{-2}$ to $10^{2}$.

\subsection{Planet Accretion and Boundary Conditions} \label{subsec:bou_con}

We follow \citet{Li2024} to implement the accretion onto the planet and the boundary condition both at the inner and outer edge of the disk. Basically, there are two parameters, which are sink hole radius $r_{\rm acc}$ around the accreting planet, a removal rate $f$ within the sink hole, to describe the dynamical accretion of the planet. 
The planetary accretion rates $\dot{m}_{\rm p}$  are recorded on-the-fly during each simulation, which are calculated as \citep[e.g.,][]{Kley2001, DAngelo2003,Li2021a,Li2023,Li2024}
\begin{equation}
\dot{m}_{\rm p} = \int_{\delta r < r_{\rm acc}} f \Sigma \ {\rm d}S,
\end{equation}
where $\Sigma$, $dS$, $\delta r$ are the surface density, surface area of each grid, and the distance to the planet, respectively.
We adopt an accretion radius as the same of the softening length with $r_{\rm acc}=0.1\ r_{\rm h}$ and a removal rate of $f=5 \Omega$, similar to \citet{Li2024}. Other accretion prescriptions have also been tested for the convergence as shown in the Appendix~\ref{subsec:con_test}.
The accreted mass and angular momentum are not added actively onto the planet, while their effect on the migration dynamics of the planet is analyzed by post-processing the gravitational and accretional torque components as in \citet{Li2024}. 

For the radial boundary, we maintain a steady inflow $\dot{m}_{\rm d}$ from the outer boundary, and preserve a radially constant inward mass flux at the inner edge as in \citet{Li2024}. A wave damping is applied at the outer and inner edges to remove the wave reflection \citep{2006MNRAS.370..529D}. A periodical boundary is adopted at the azimuthal direction.

\section{SIMULATION RESULTS} \label{sec:sim_results}

We first outline the accretion dynamics of the planet, followed by an examination of the effects of planetary migration under various cooling prescriptions.

\subsection{The Accretion of the Planet} \label{subsec:re_acc}

\begin{figure}
\begin{adjustwidth}{-\extralength}{0cm}
\centering
\includegraphics[width=0.55\textwidth]{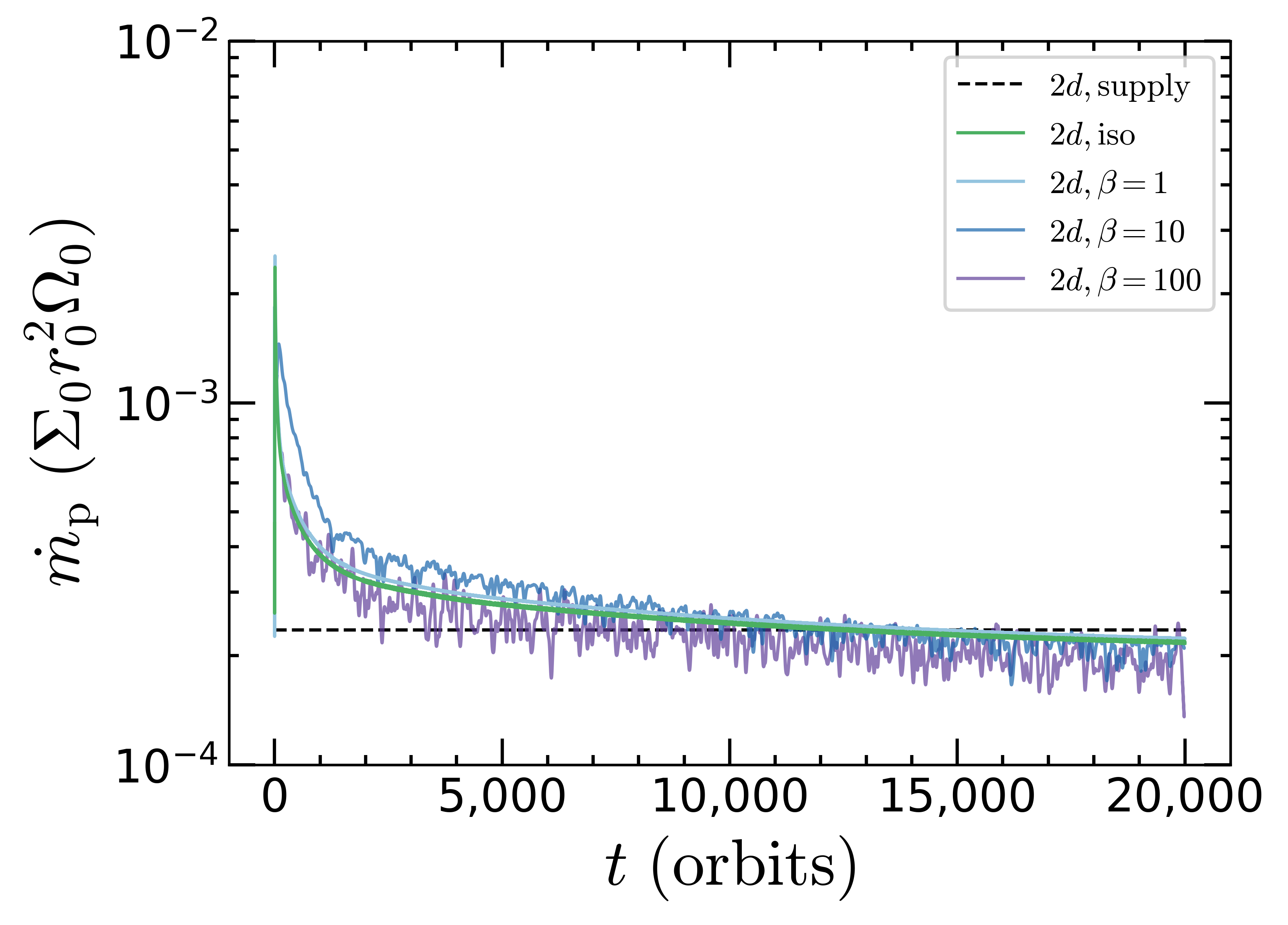}
\includegraphics[width=0.55\textwidth]{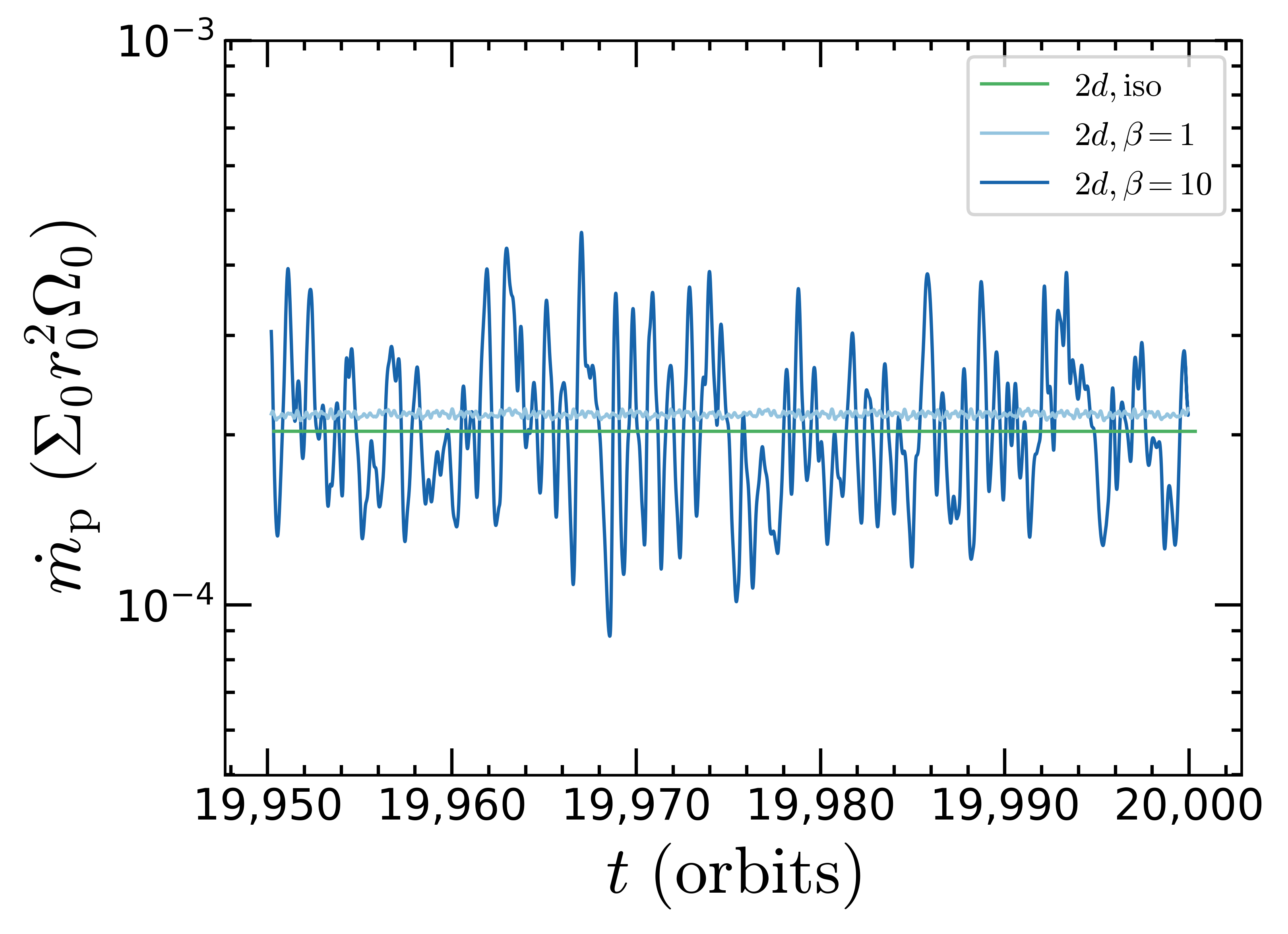}
\end{adjustwidth}
\caption{The left panel displays the running time-averaged planetary accretion rates in a locally isothermal disk and disks with varying $\beta$-cooling timescales, measured in unit of $\Sigma_{0} r_{0}^{2} \Omega_{0}$. The black dashed line indicates the material supply rate from the outer boundary. The right panel illustrates the short-term fluctuations in accretion rates for several representative simulations, without any time averaging.}
\label{fig:accrate_2d}
\end{figure}

The time evolution of the planetary accretion rates $\dot{m}_{\rm p}$ for isothermal and different $\beta-$cooling are shown in Figure~\ref{fig:accrate_2d}. 
All the simulations proceed to 20000 orbits, which is roughly the viscous timescales at $r_{\rm out}$ (one viscous timescale can be calculated as $(2\pi \alpha h_{0}^{2})^{-1}$ orbits). 
The accretion rates at 20000 orbit for different runs are slightly below the mass supply rate $\dot{m}_{\rm d}=3\pi\Sigma\nu$ at the outer boundary, similar to \citep{Li2024}. This suggests that most of the mass supply from the outer disk is accreted onto the planet, with a small fraction of the disk supply accreted onto the central star. The nearly constant accretion rates onto the planet around 20000 orbits indicates the viscous quasi-steady state for the CPD accretion. We have further checked the mass flux across the Hill sphere around the planet, which is nearly constant as a function of $\delta r$ and in good agreement with the accretion rates onto the planet. The accretion rate also matches well with the global disk mass flux jump at the the planetary orbit. All of these confirm that the quasi-steady state for the CPD accretion is well established. 
For different cooling prescriptions, it can be seen that a higher $\beta$ results in a slightly decrease in the planetary accretion rate $\dot{m}_{\rm p}$, although strong variabilities appear for high $\beta$-cooling runs.

\begin{figure*}
\begin{adjustwidth}{-\extralength}{0cm}
\centering
\includegraphics[width=0.4\textwidth]{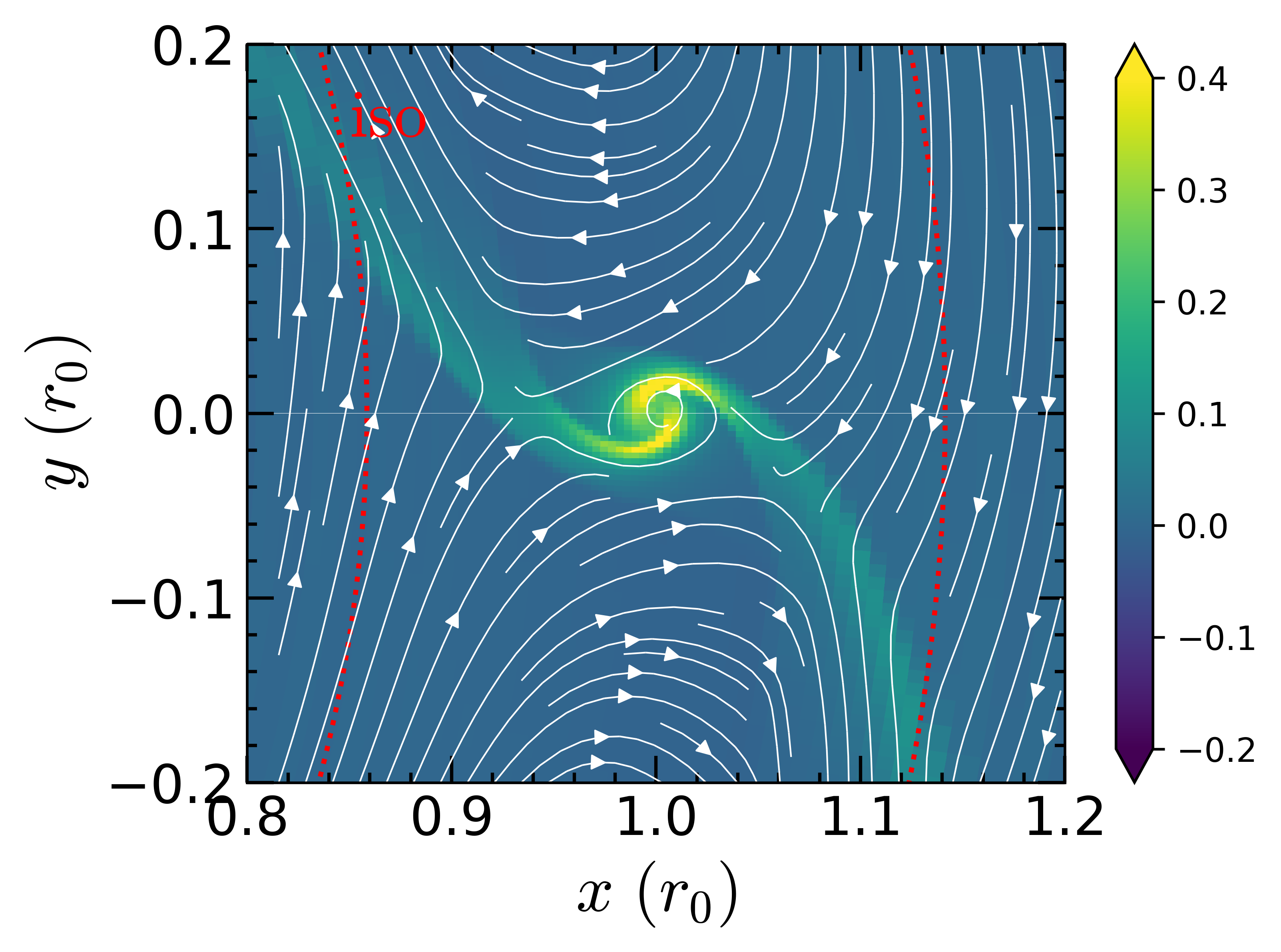}
\includegraphics[width=0.4\textwidth]{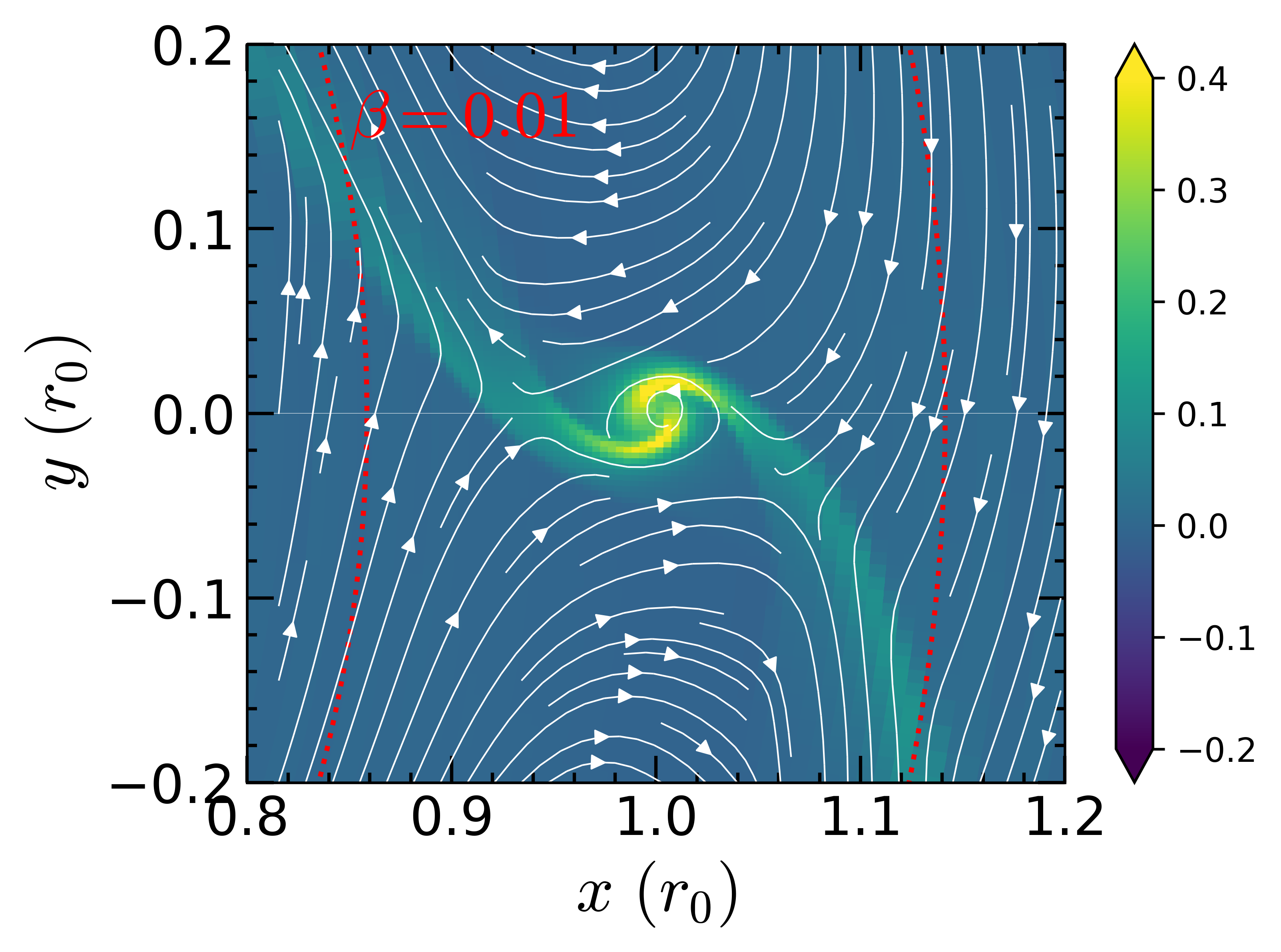}
\includegraphics[width=0.4\textwidth]{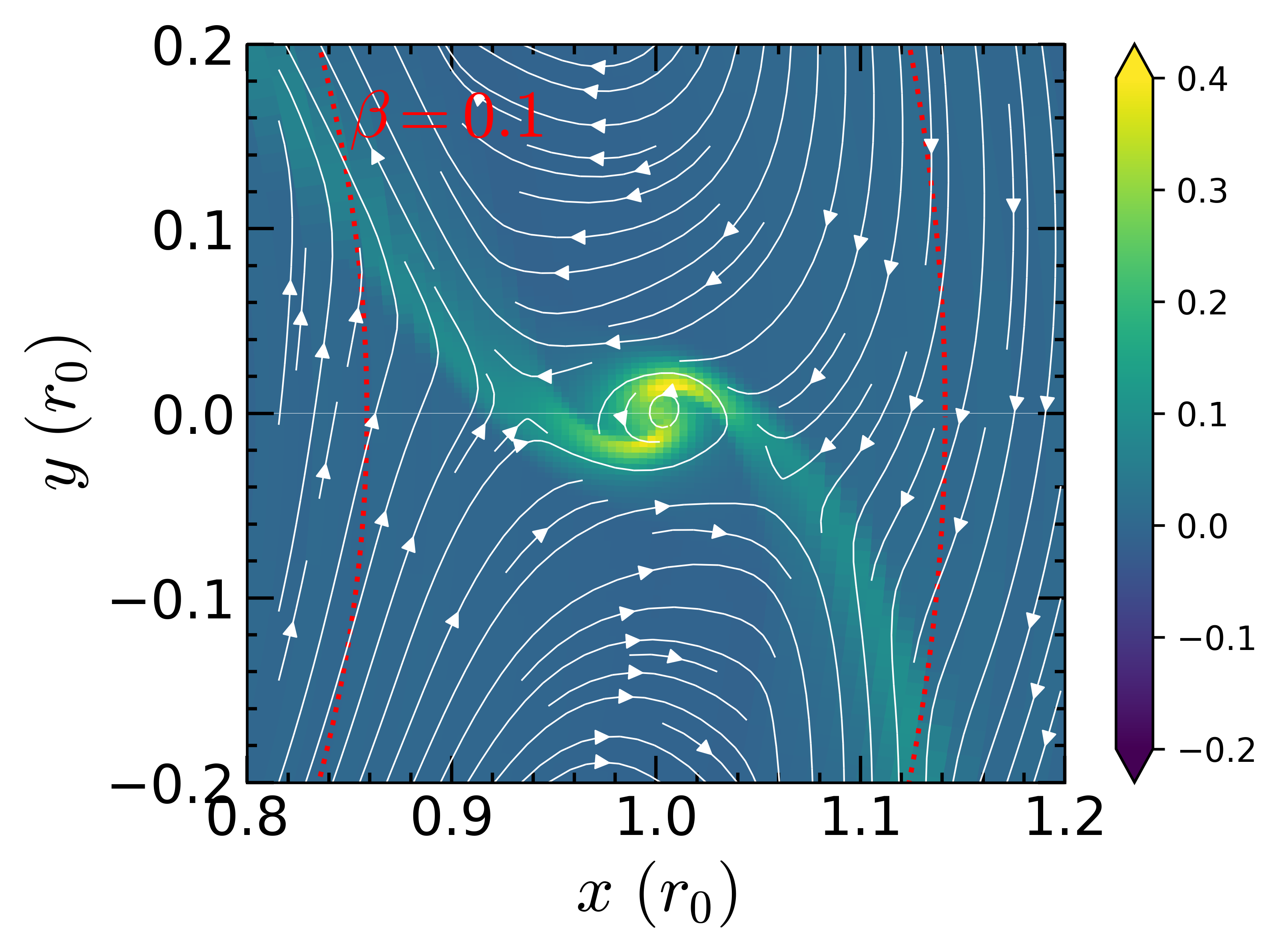}

\includegraphics[width=0.4\textwidth]{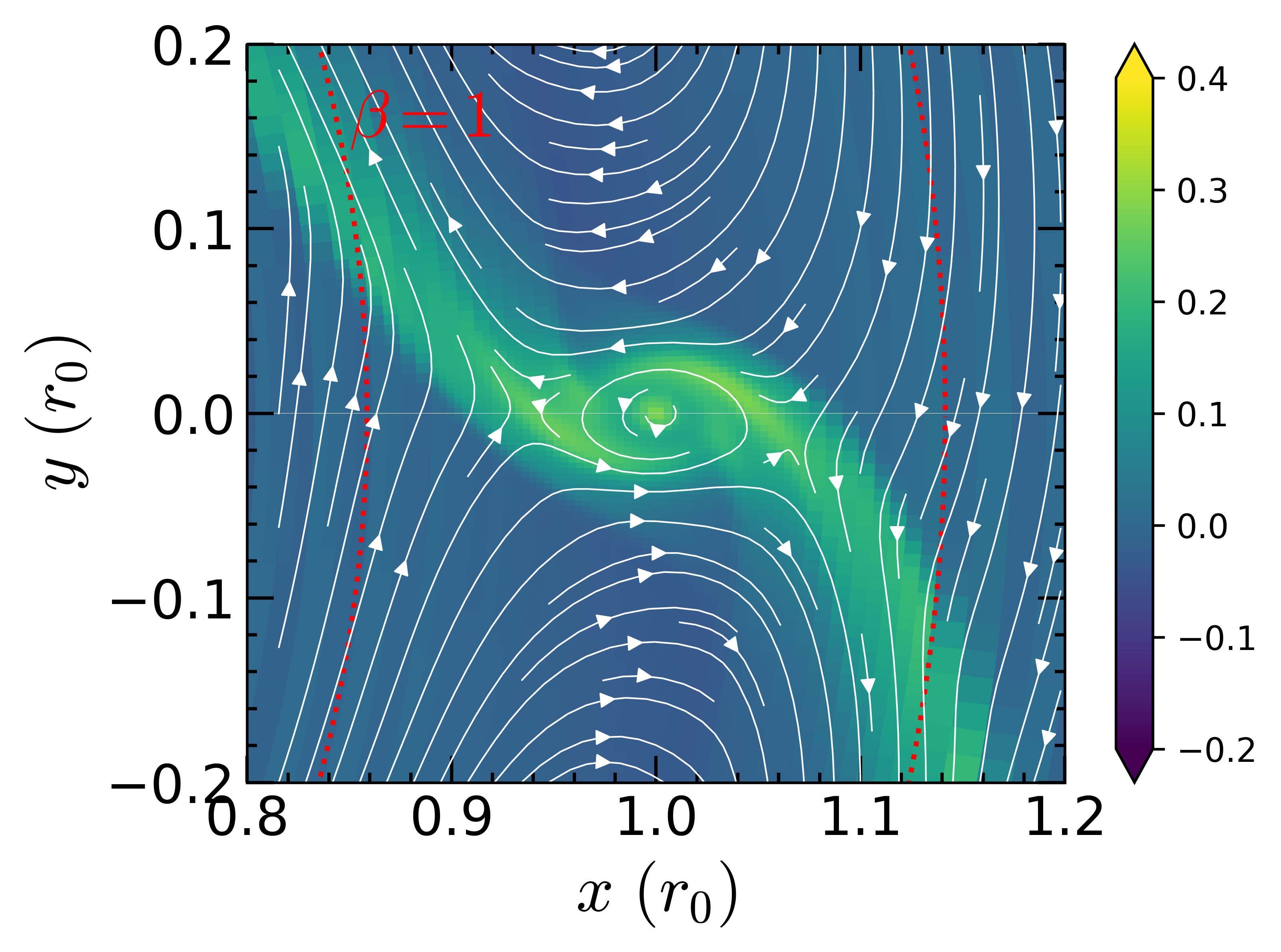}
\includegraphics[width=0.4\textwidth]{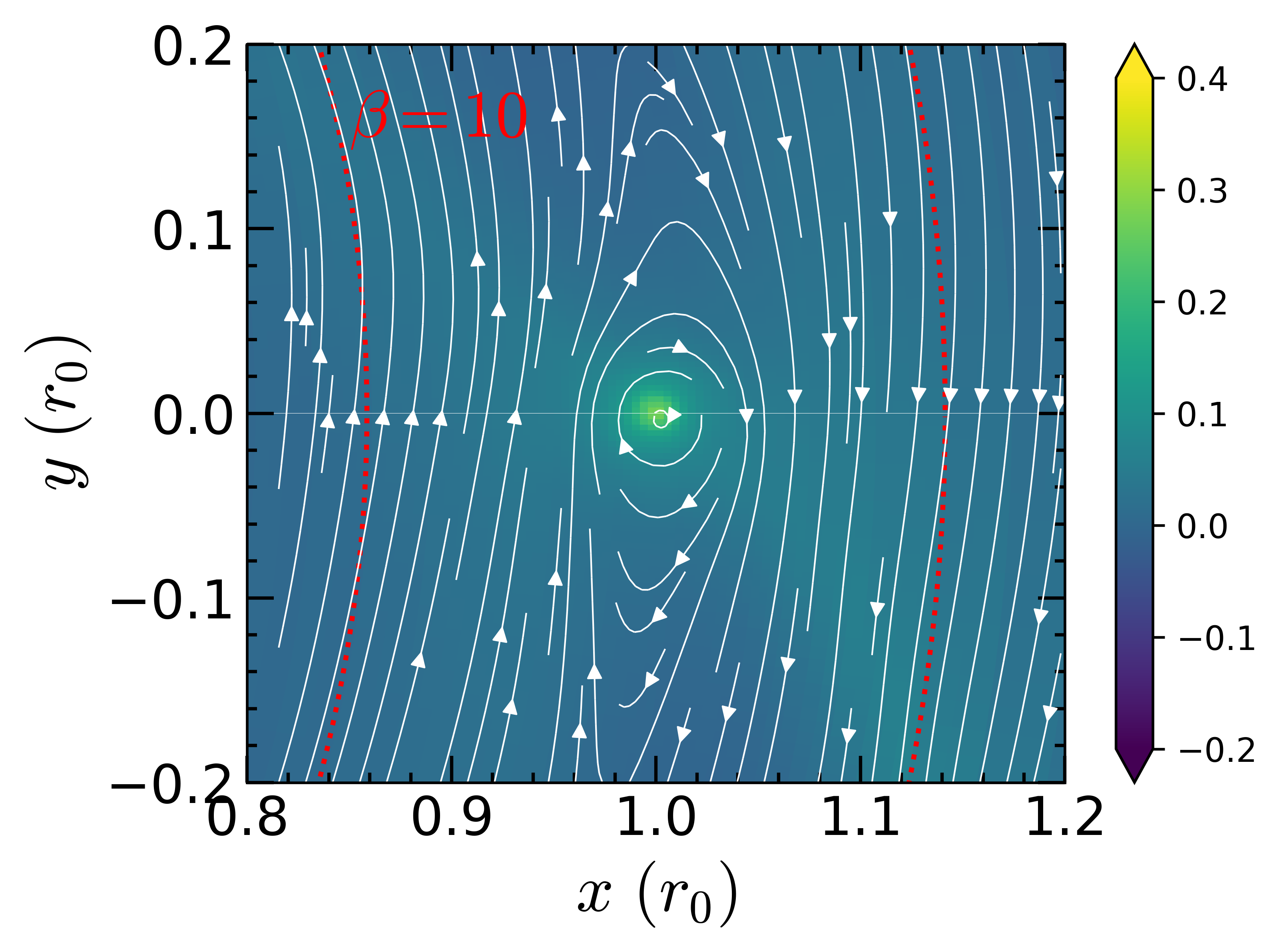}
\includegraphics[width=0.4\textwidth]{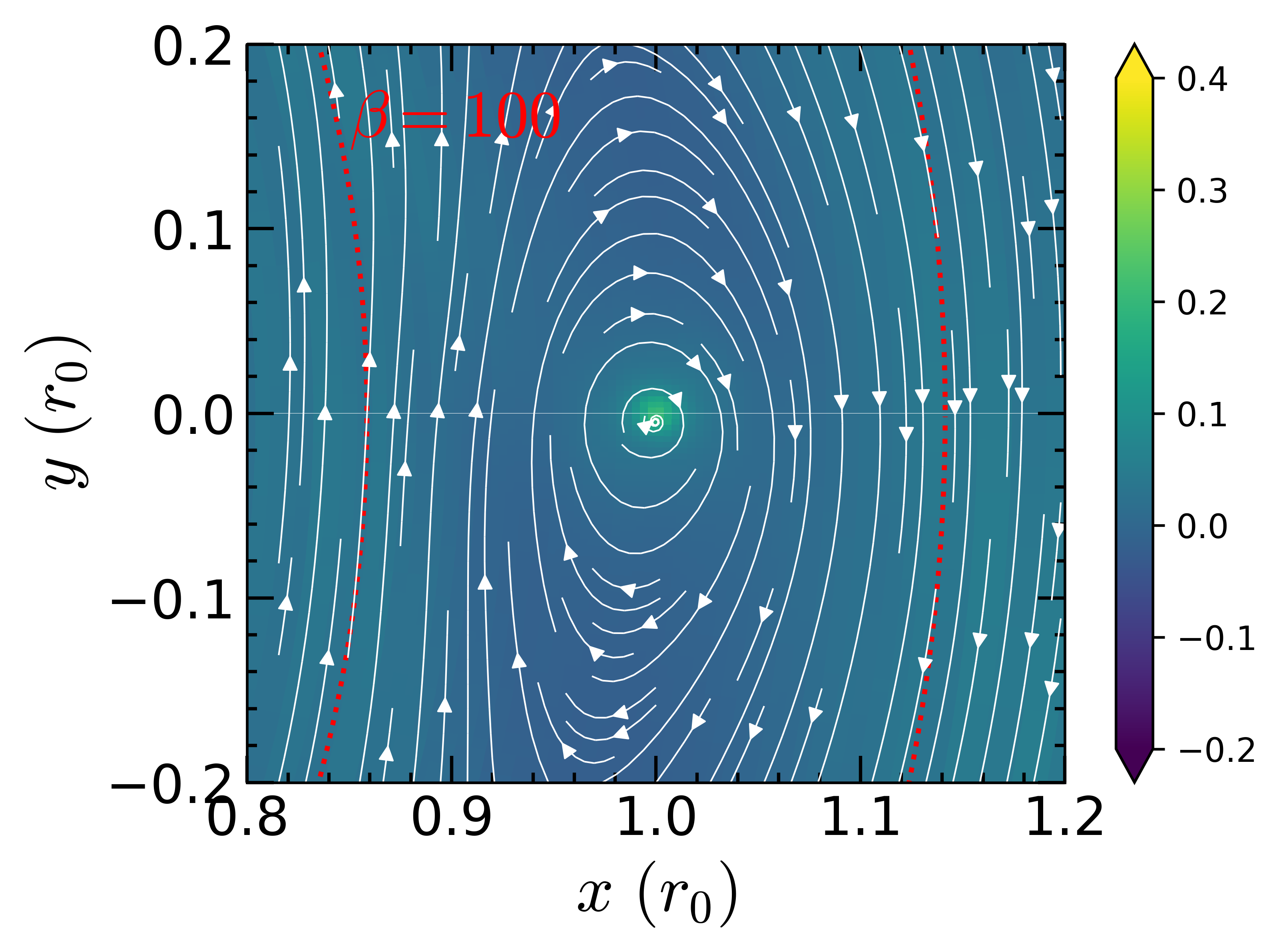}
\end{adjustwidth}
\caption{The density distribution $\delta \Sigma$ in the vicinity of the planet with different cooling timescales. $\delta \Sigma = \Sigma - \langle \Sigma \rangle$ where $\langle \Sigma \rangle$ is the azimuthally averaged surface density from the perturbed disk at the same orbital time. The arrows in the plots represent the velocity vectors relative to the planet. All plots are presented at 20000 orbits. For the cases with $\beta=10$ and $\beta=100$, time averages over 100 and 500 orbits, respectively, were performed to achieve steady flow structures. The red dotted lines enclose the classical horseshoe region around the planetary orbit defined as $x_{\rm s}=a_{\rm p}(q/h)^{1/2}$ \citep[e.g.,][]{Kley2012}.}
\label{fig:CPD_2d}
\end{figure*}

The surface density distribution and the velocity field in the frame corotating with the planet are shown in Figure \ref{fig:CPD_2d}. The gas velocity field reveals the presence of a distinct disk structure in close proximity to the planet, known as the CPD, in the isothermal case. The disk structure is particularly prominent under isothermal condition. Additionally, the horseshoe streamline is clearly visible for the isothermal and low-$\beta$ cases, but it disappears for $\beta\gtrsim10$. It is evident that the accreted material is brought into the vicinity of the planet from the upper horseshoe orbits and flows outward from the spiral density waves. Notably,  the rotational velocity is in the retrograde direction for $\beta\gtrsim10$, in contrast to the prograde flow observed in the low $\beta$ cases.

\begin{figure*}
\begin{adjustwidth}{-\extralength}{0cm}
\centering
\includegraphics[width=0.4\textwidth]{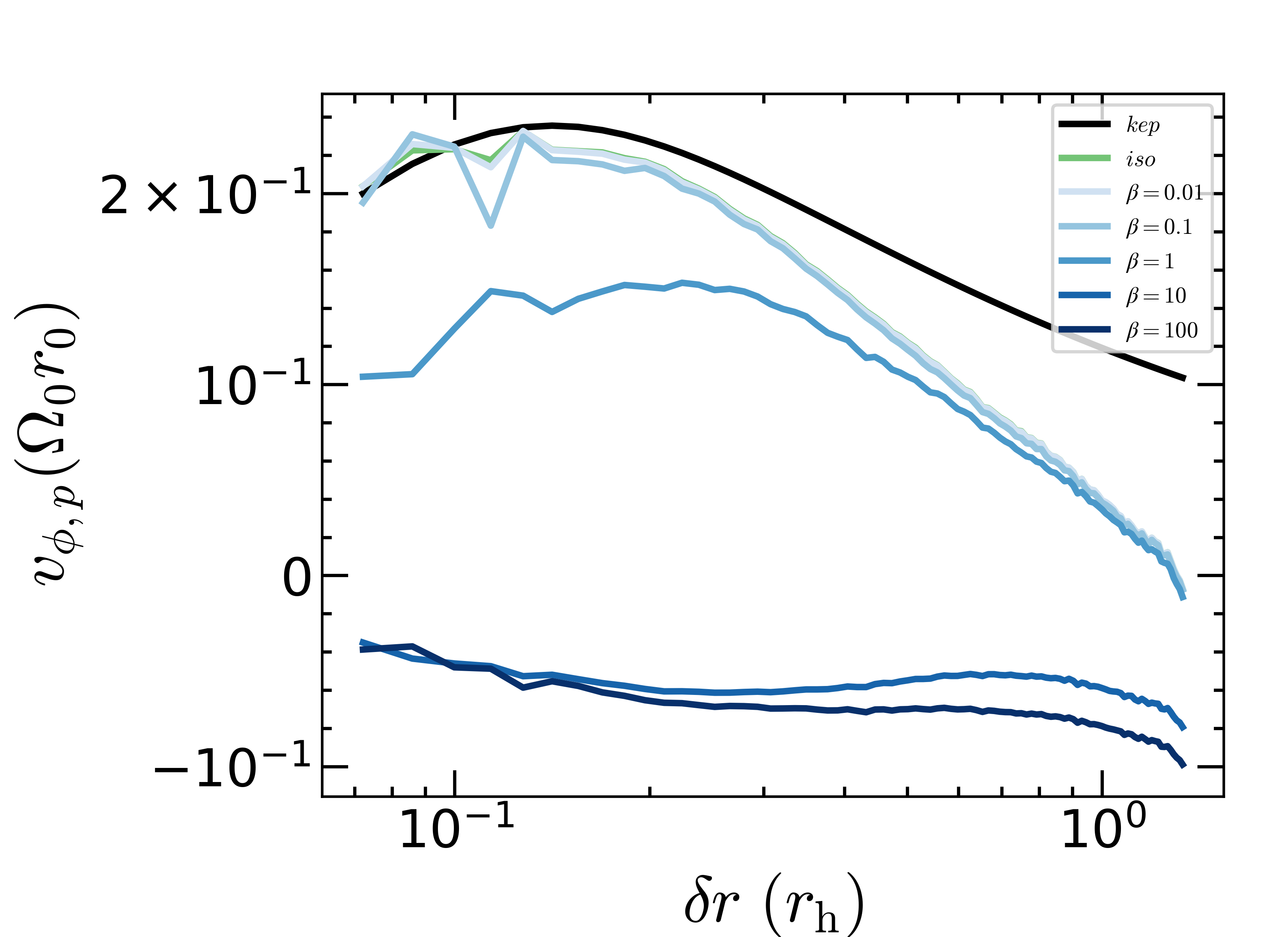}
\includegraphics[width=0.4\textwidth]{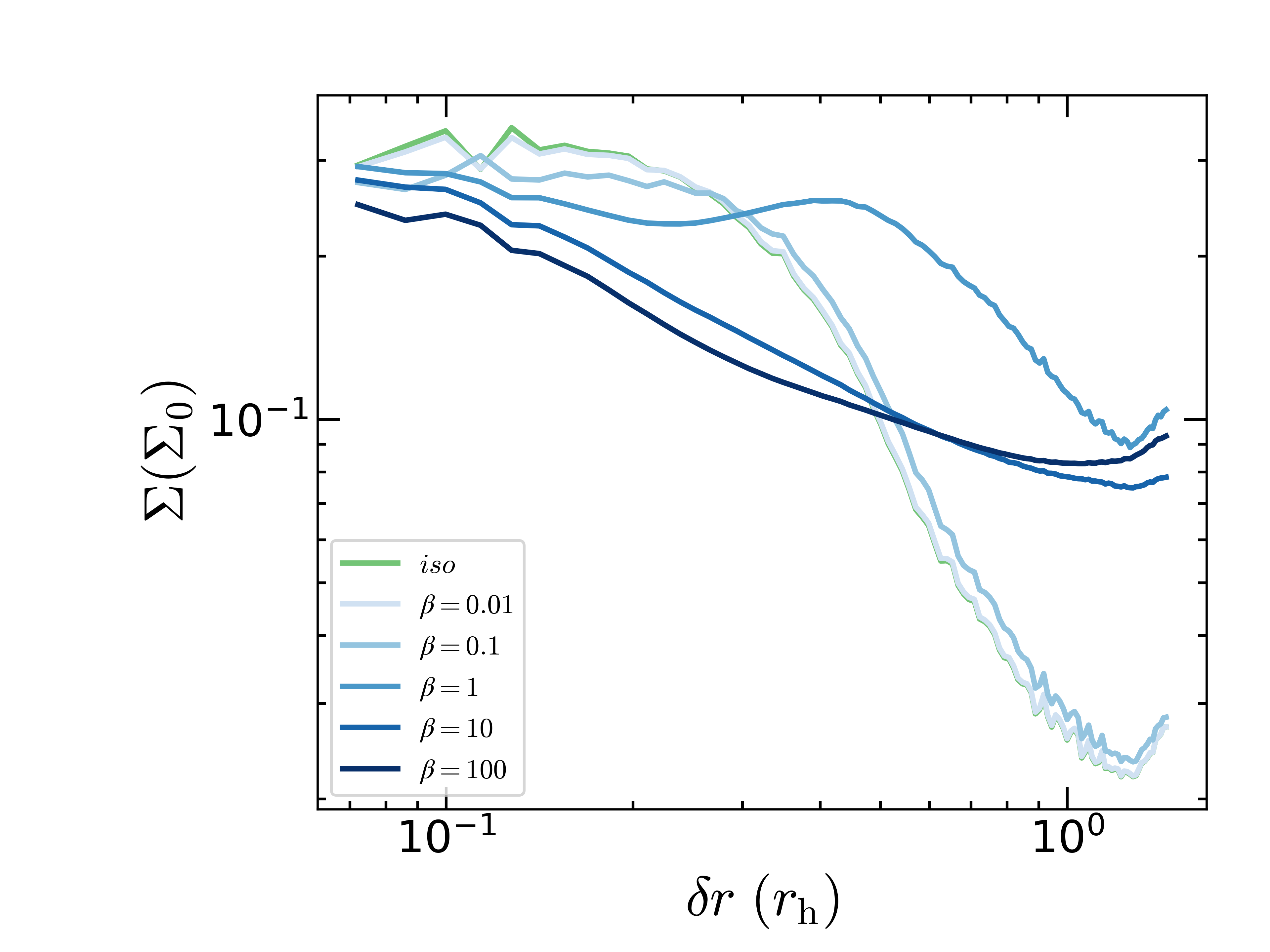}
\includegraphics[width=0.4\textwidth]{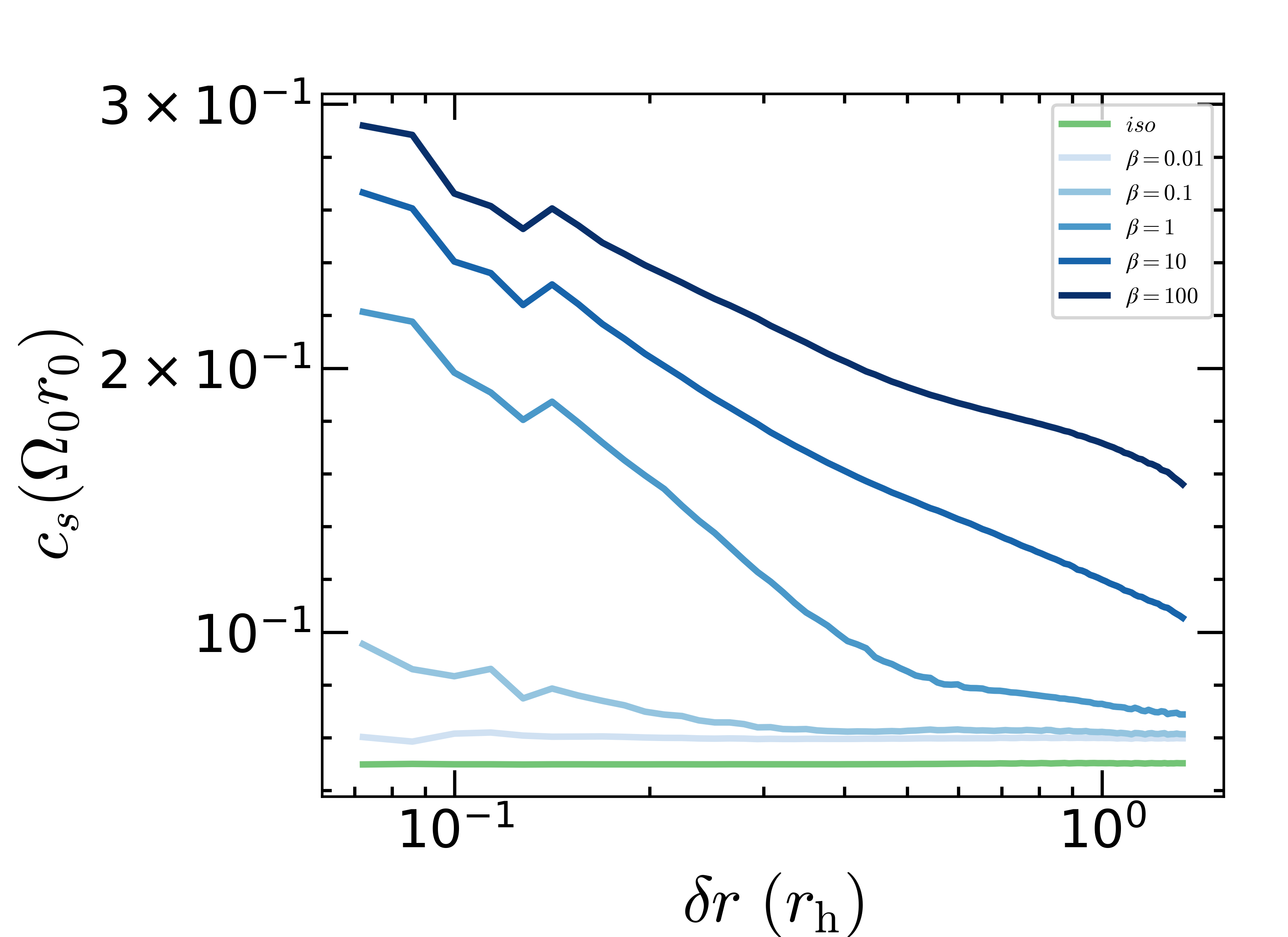}
\end{adjustwidth}
\caption{Left panel: the azimuthally-averaged rotation velocity relative to the planet for different models in unit of the local Keplerian velocity orbiting around the star. The black line represents the Keplerian rotation velocity around the planet. Middle panel: the surface density profile around the planet in unit of $\Sigma_{0}$. Right panel: the azimuthally-averaged sound speed $c_{\rm s}$ in the vicinity of the planet in code unit. 
}
\label{fig:CPD_1d}
\end{figure*}

To facilitate a more comprehensive examination of the CPD structures in isothermal and long cooling timescale cases, we plot the azimuthally-averaged rotation curves and density profiles of the CPDs in Figure \ref{fig:CPD_1d}. 
Based on the rotation curves presented in the left panel, it is evident that the rotation velocity within $\sim0.2\ r_{\rm h}$ under the isothermal condition closely resembles the Keplerian rotation. This is also true for the cases with $\beta\lesssim 0.1$. This indicates that the CPD is rotationally supported up to $\sim0.2\ r_{\rm h}$ and exhibits characteristics of a disk structure in these cases, where the centrifugal force effectively counterbalances the gravitational force. In the longer cooling timescale cases, the rotation curve exhibits a significant deviation from the Keplerian curve in all regions within the Hill sphere and the materials even rotate in the retrograde direction for $\beta\gtrsim10$ as confirmed in Figure~\ref{fig:CPD_2d}. 
Consequently, it is the pressure gradient due to the inefficient cooling in the CPD region that balances the gravity.
This is consistent with previous 3D simulations which have shown that the rotationally supported CPD appears mostly under the isothermal EOS, while it transformed into a pressure supported envelope for an adiabatic EOS \citep[e.g.,][]{Fung2019}. 
Actually, based on the temperature (sound speed) profile within the Hill sphere shown in the right panel of Figure~\ref{fig:CPD_2d}, the Bondi radius $r_{\rm Bondi}\equiv m_{\rm p}/c_{\rm s}^{2}$ at the expected CPD size from the planet (e.g., $\simeq0.2\ r_{\rm h}$) is smaller than the Hill radius due to the increase of the temperature in the vicinity of the planet, which suggests that the planetary accretion proceeds indeed by the Bondi-like process.

\subsection{Mechanism of the Accretion} \label{subsec:dis_acc}

To explore the accretion process, we examine the CPD's angular momentum conservation equation in the vicinity of a planet following \citet{Zhu2016}, 

\begin{equation}\label{equ:ang_mom}
\partial_{t}\langle \Sigma \delta rv_{\phi}^{\prime}\rangle_{\phi} = -\frac{1}{\delta r}\partial_{r}(\delta r^2\langle{\Sigma}v_{r}^{\prime}v_{\phi}^{\prime}\rangle_{\phi}) + \langle\vec{\delta r}\times\vec{F}\rangle_{\phi},
\end{equation}
where $\delta r$ is the radial distance relative to the planet and $\langle{X}\rangle_{\phi}$ represents $\int_{-\pi}^{\pi} X {\rm d}\phi$. The quantities with $^{\prime}$ indicate that they are measured in the frame relative to the planet. Here we have neglected the viscous stress associated with the explicit alpha viscosity in the CPD since it is found that the explicit $\alpha$ viscosity is much smaller than the effective viscosity due to shock dissipation. The sum of all external forces, denoted as $\vec{F}$, encompasses various components, including stellar gravity, centrifugal forces and Coriolis forces within the reference frame that orbits the star with the planet's orbital period. The planet's gravity does not contribute to the external forces because it is in parallel to the radial vector $\vec{\delta r}$.

By decomposing the velocity in $\phi$ direction into two components, the averaged velocity $\overline{v^{\prime}}$ (it is worth noting that we use the azimuthally averaged velocity in the $\phi$ direction in instead of the local Keplerian velocity due to the distinct deviation from Keplerian motion under long cooling timescales) and the disturbance velocity $\delta v_{\phi}^{\prime}$, and afterwards multiplying both sides of Equation~(\ref{equ:ang_mom}) by $\delta r^{2}$ to ensure angular momentum dimensions, the equation below can be obtained,

\begin{equation}\label{equ:Ang_mom_t}
\delta r^2\partial_{t}\langle \Sigma \delta rv_{\phi}^{\prime}\rangle_{\phi} = \dot{M}\delta r\partial_{r}(\delta r \overline{v^{\prime}}) - \delta r\partial_{r}(\delta r^2\langle{\Sigma}v_{r}^{\prime }v_{\phi}^{\prime}\rangle_{\phi})
+ \delta r^2\langle\vec{\delta r}\times\vec{F}\rangle_{\phi},
\end{equation}
where $\dot{M}_{\rm CPD}=-\left<\Sigma \delta rv_{r}^{\prime}\right>_{\phi}$. It turns out the mass flux in the CPD $\dot{M}_{\rm CPD}$ well matches with the planetary accretion rate $\dot{m}_{\rm p}$, which again suggests a steady state for the CPD accretion. The term on the left side represents the time-dependent evolution term of angular momentum, the first term on the right side denotes the angular momentum transfer resulting from accretion, the second term corresponds to the gradient of Reynolds stress or the gradient of angular momentum transfer flow, and the last term represents the external torque.

When the state of accretion equilibrium for the CPD has been established as in our cases, the time-dependent evolution term in Equation~(\ref{equ:Ang_mom_t}) is negligible. Consequently, Equation~(\ref{equ:Ang_mom_t}) can be expressed as follows,

\begin{equation} \label{equ:Ang_mom}
-\dot{M}_{\rm CPD}\delta r\partial_{r}(\delta r \overline{v^{\prime}})\approx -\delta r\partial_{r}(\delta r^2\langle{\Sigma}v_{r}^{\prime}v_{\phi}^{\prime}\rangle_{\phi}) + \delta r^2\langle\vec{\delta r}\times\vec{F}\rangle_{\phi}.
\end{equation}
By performing integration on both sides of Equation~(\ref{equ:Ang_mom}), the components of the aforementioned equation is written as,

\begin{equation} \label{equ:alpha}
\alpha_{\rm eff} = \alpha_{\rm rey} + \alpha_{\rm con} + \alpha_{\rm T}, 
\end{equation}
where 

\begin{equation} \label{equ:alpha_eff}
\alpha_{\rm eff} = \frac{\dot{M}_{\rm CPD}}{3 \pi \Sigma c_{\rm s} H},
\end{equation}

\begin{equation} \label{equ:alpha_Rey}
\alpha_{\rm rey} = \frac{\langle{\Sigma}v_{r}^{\prime} \delta v_{\phi}^{\prime}\rangle_{\phi}}{3 \pi \Sigma c_{\rm s}^{2}},
\end{equation}

\begin{equation} \label{equ:alpha_con}
\alpha_{\rm con} = \frac{C}{3 \pi \Sigma c_{\rm s} H \delta r \overline{v^{\prime}}},
\end{equation}

\begin{equation} \label{equ:alpha_T}
\alpha_{\rm T} = -\frac{\int \delta r \langle\vec{\delta r} \times \vec{F}\rangle_{\phi}dr}{3 \pi \Sigma c_{\rm s} H \delta r \overline{v^{\prime}}},
\end{equation}
where $\alpha_{\rm eff}$ represents the effective accretion coefficient of the CPD, while $\alpha_{\rm rey}$ denotes the accretion coefficient resulting from Reynolds stress, $\alpha_{\rm T}$ is the accretion coefficient induced by external force. $\alpha_{\rm con}$ is associated with the accretion coefficient determined by a constant free parameter $C$ associated with the accretion boundary of the CPD.  

\begin{figure}
\begin{adjustwidth}{-\extralength}{0cm}
\centering
\includegraphics[width=0.55\textwidth]{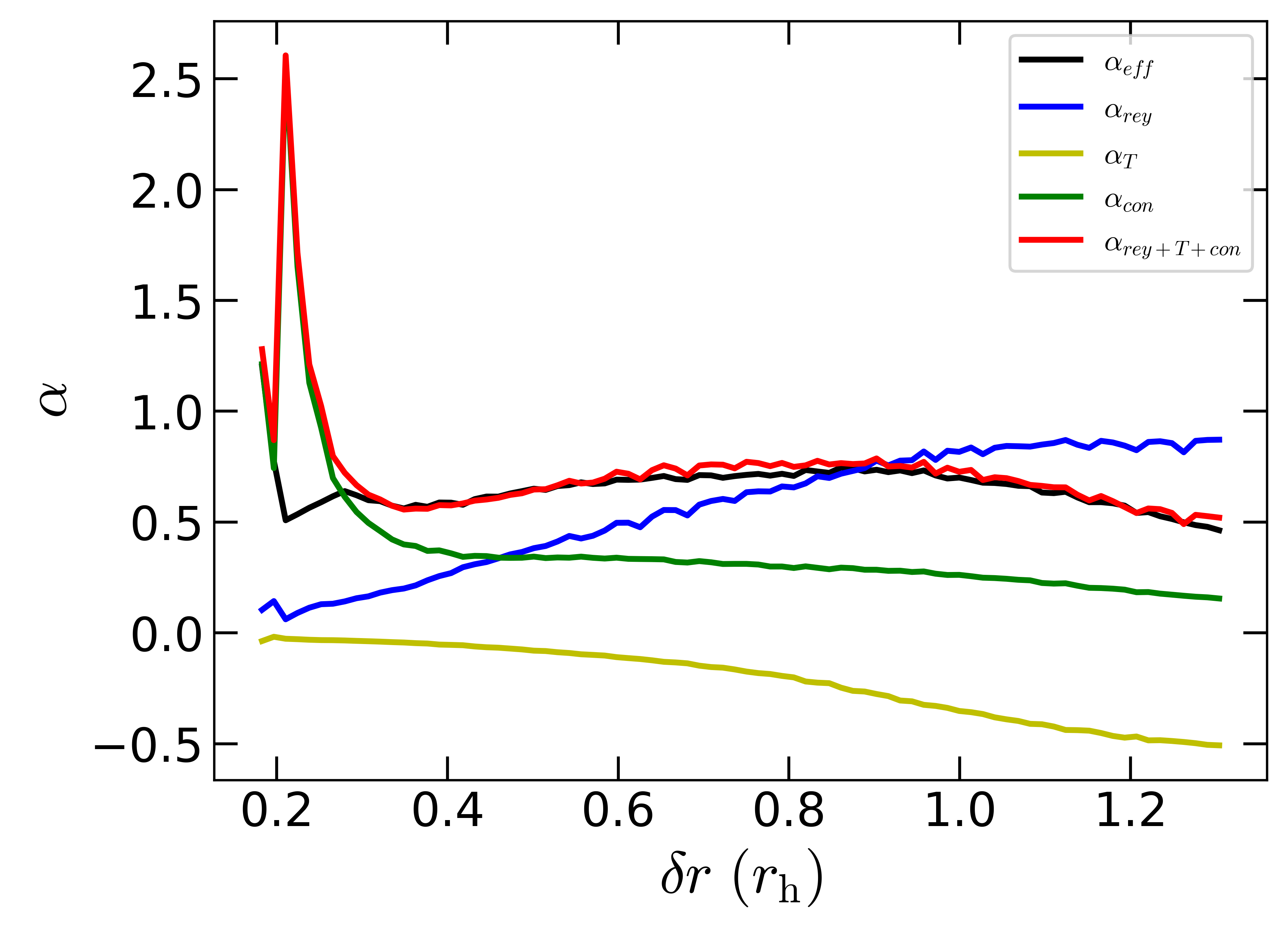}
\includegraphics[width=0.55\textwidth]{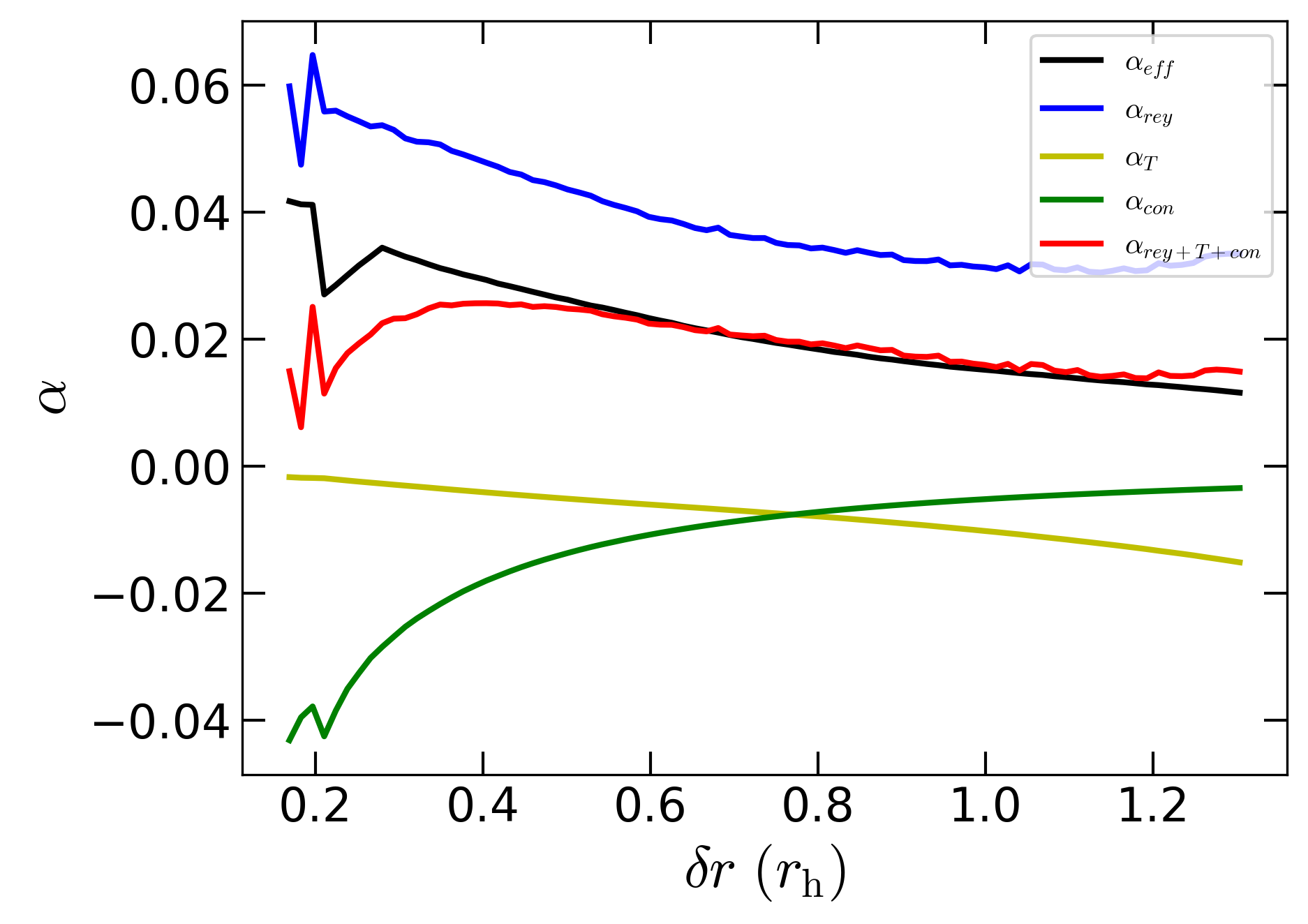}
\end{adjustwidth}
\caption{Different accretion coefficient calculated by Equation \ref{equ:alpha} under isothermal EOS (left panel) and $\beta=100$ (right panel) cases. The the summation of $\alpha_{\rm rey}$, $\alpha_{\rm con}$ and $\alpha_{\rm T}$ is shown as red lines.
}
\label{fig:alpha_iso}
\end{figure}

As shown in Figure \ref{fig:alpha_iso}, it is evident that the total accretion coefficient ($\alpha_{\rm rey}+\alpha_{\rm T}+\alpha_{\rm con}$)  by combing different components can be in reasonable agreement  with the effective $\alpha_{\rm eff}$ based on measured planetary accretion rate $\dot{m}_{\rm p}$ for both isothermal and $\beta=100$ cases. It should be noted that due to the softening lengths and the inner accreting boundary in the vicinity of the planet, the two curves do not align perfectly.  
Therefore, the consistency of Equation~(\ref{equ:alpha}) is another signature of a steady state within the CPD region.
Furthermore, we can see that $\alpha_{\rm eff}$ is quite close to $\alpha_{\rm rey}$ in most parts of the disk, and both $\alpha_{\rm eff}$ and $\alpha_{\rm rey}$ is much higher than the explicit $\alpha$ viscosity implemented in the global disk.
These suggest that the angular momentum transport and thus the accretion within the CPD could be mainly induced by spiral shocks in the CPD regions instead of the global disk viscosity \citep{Zhu2016}.

\begin{figure}
\begin{adjustwidth}{-\extralength}{0cm}
\centering
\includegraphics[width=0.55\textwidth]{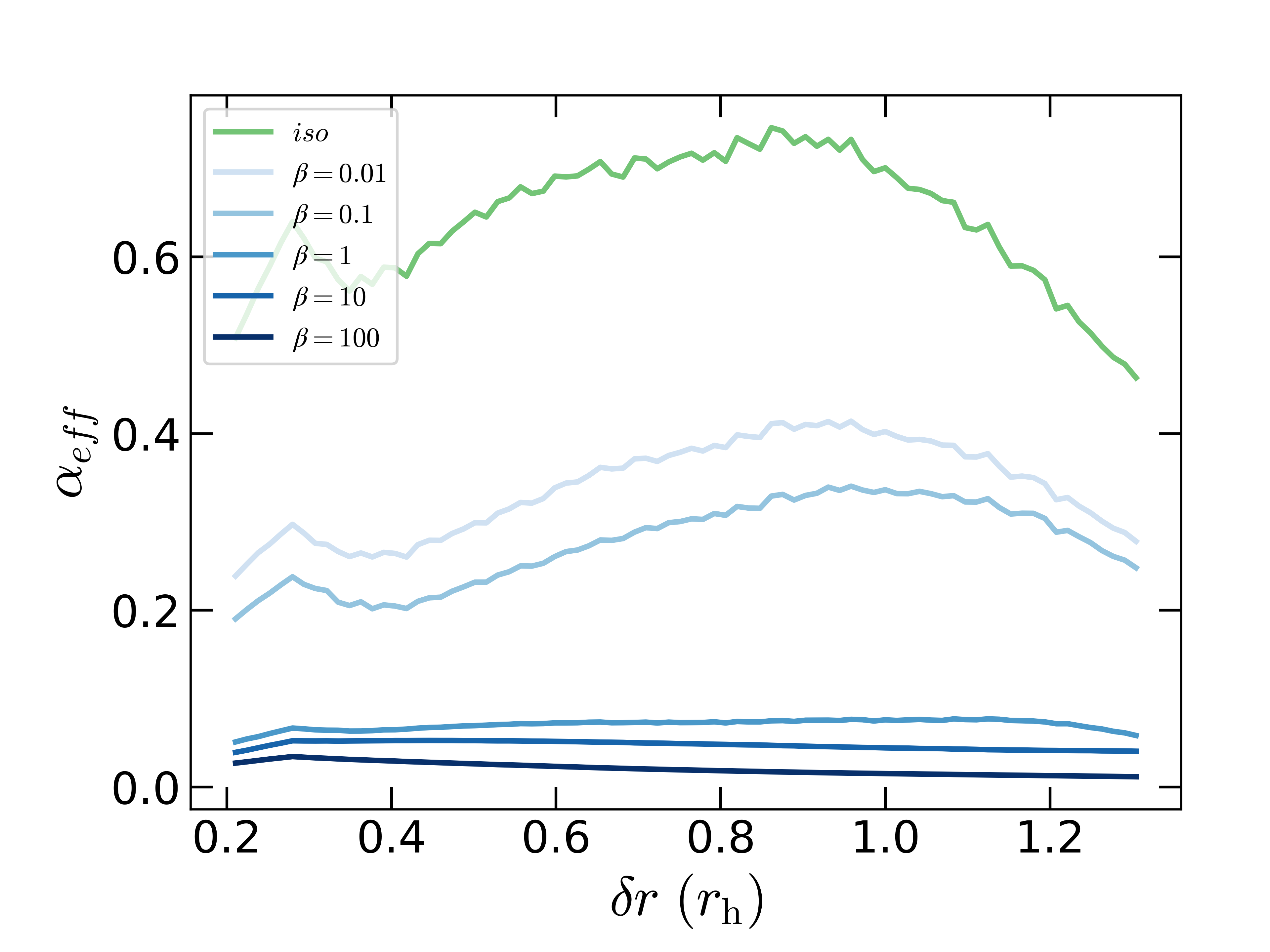}
\includegraphics[width=0.55\textwidth]{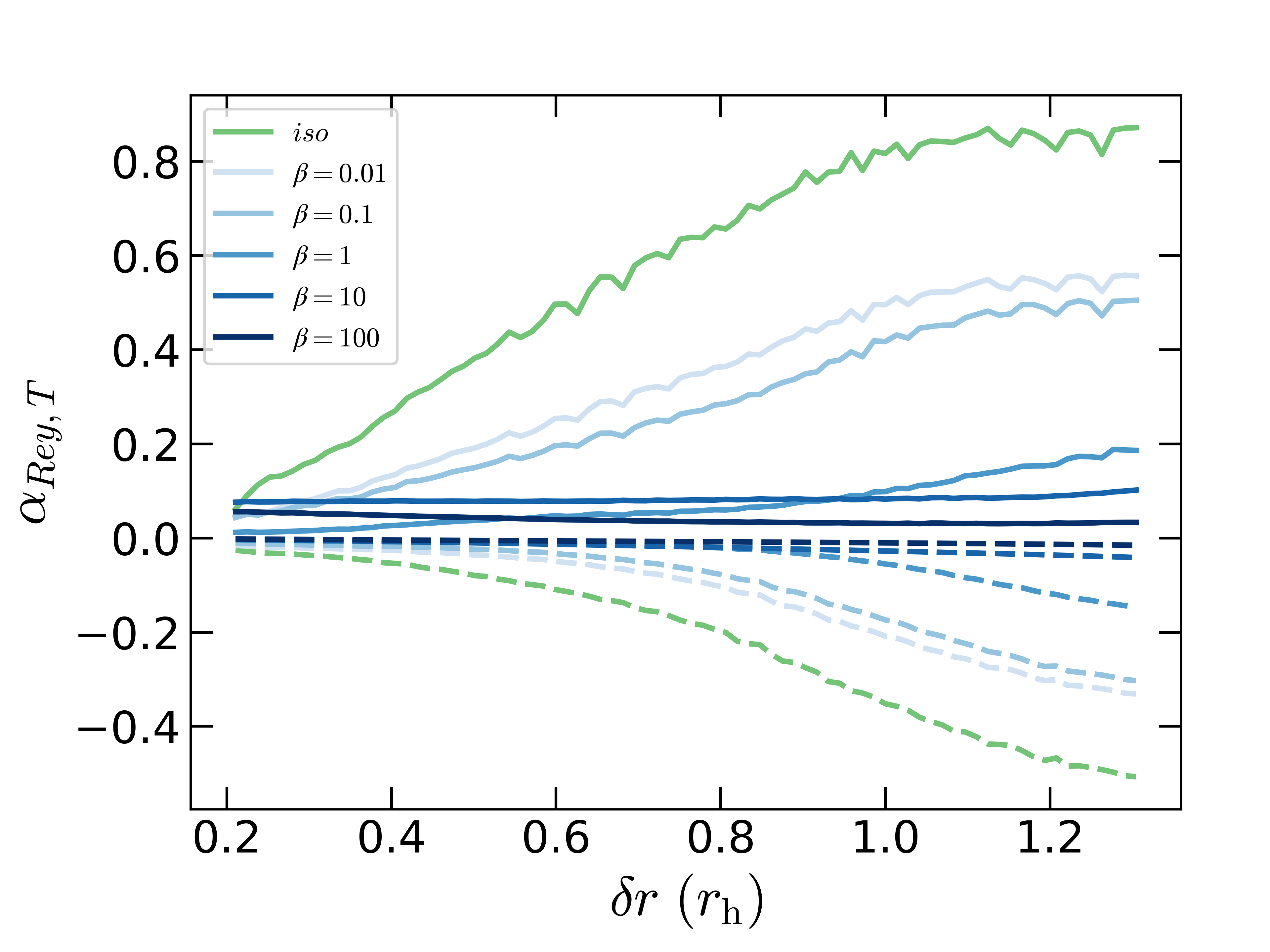}
\end{adjustwidth}
\caption{Left panel: the effective accretion coefficients in Equation~\ref{equ:alpha_eff} under different cooling timescales. Right panel: the accretion coefficients resulting from Reynolds stress($\alpha_{\rm rey}$; solid lines) and external forces ($\alpha_{\rm T}$; dashed lines) for different cooling timescales. The same color represents the same cooling timescale. 
}
\label{fig:alpha_2d}
\end{figure}

The effective accretion coefficients $\alpha_{\rm eff}$ for different $\beta$-cooling models are shown in the left panel of Figure \ref{fig:alpha_2d}, illustrating its variation across various thermodynamic factors. It reveals that the effective accretion coefficient is highest under the isothermal condition and progressively decreases as the cooling timescale increases, ultimately reaching its minimum under adiabatic conditions. As the net accretion rates onto the planet vary slightly for different $\beta$-cooling, the decreasing of $\alpha_{\rm eff}$ is partly compensated by a higher CPD temperature resulting from the adiabatic heating with a longer cooling timescale, as shown in the right panel of Figure~\ref{fig:CPD_1d}. 
Nevertheless, the mass flux across the CPD and the accretion rate onto the planet under the isothermal limit are still the highest one after considering the lowest disk temperature for allowing the formation of a rotationally supported disk. 

Furthermore, the Reynolds stress coefficient $\alpha_{\rm rey}$ and the external torque coefficient $\alpha_{\rm T}$ under different cooling timescales are represented in the right panel of Figure \ref{fig:alpha_2d}. It is evident that the $\alpha_{\rm T}$ is anti-correlated with the effective accretion coefficient $\alpha_{\rm eff}$ among different thermodynamical models. The Reynolds stress coefficient $\alpha_{\rm rey}$, however, exhibits noticeable positive correlation with $\alpha_{\rm eff}$ for different cooling timescales, with the highest values observed under isothermal conditions. It, therefore, again suggests that the CPD accretion for different thermodynamic models is mostly attribute to the Reynolds stress.

In the long cooling timescale case, the gas is unable to efficiently dissipate the compressional heat, leading to an elevated temperature in the vicinity of the planet, as shown in the right panel of Figure~\ref{fig:CPD_1d}. The sound speed appeared in Equation \ref{equ:alpha_Rey} is larger compared to the isothermal limit. In addition, as the azimuthal motion relative to the planet is significantly suppressed with a longer cooling timescale, the Reynolds stress in Equation \ref{equ:alpha_Rey} also becomes weaker.
As a result, the value of $\alpha_{\rm rey}$ is reduced with the increasing of the cooling timescale, and the shock-driven accretion becomes less efficient.

\begin{figure}
\begin{adjustwidth}{-\extralength}{0cm}
\centering
\includegraphics[width=0.55\textwidth]{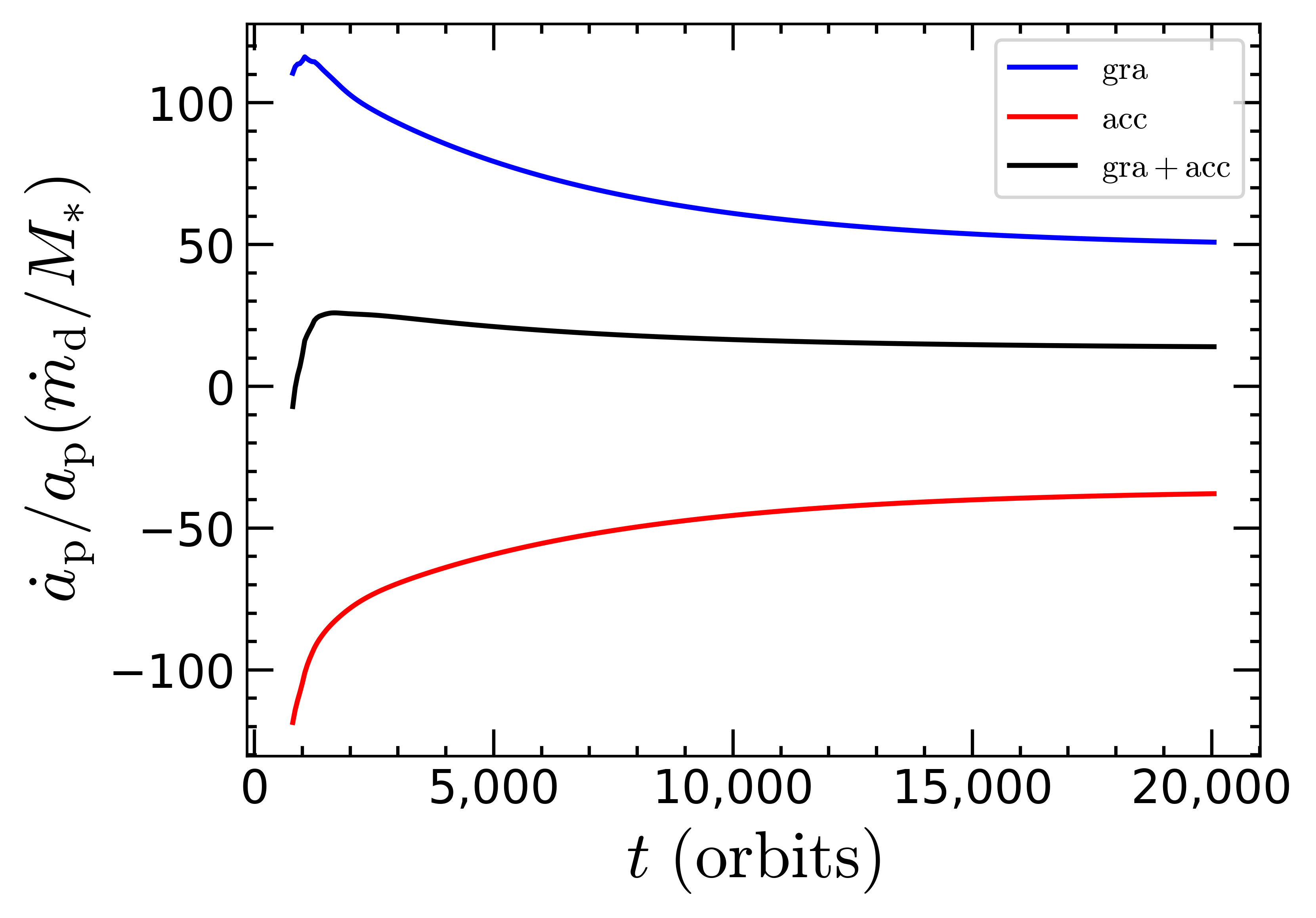}
\includegraphics[width=0.55\textwidth]{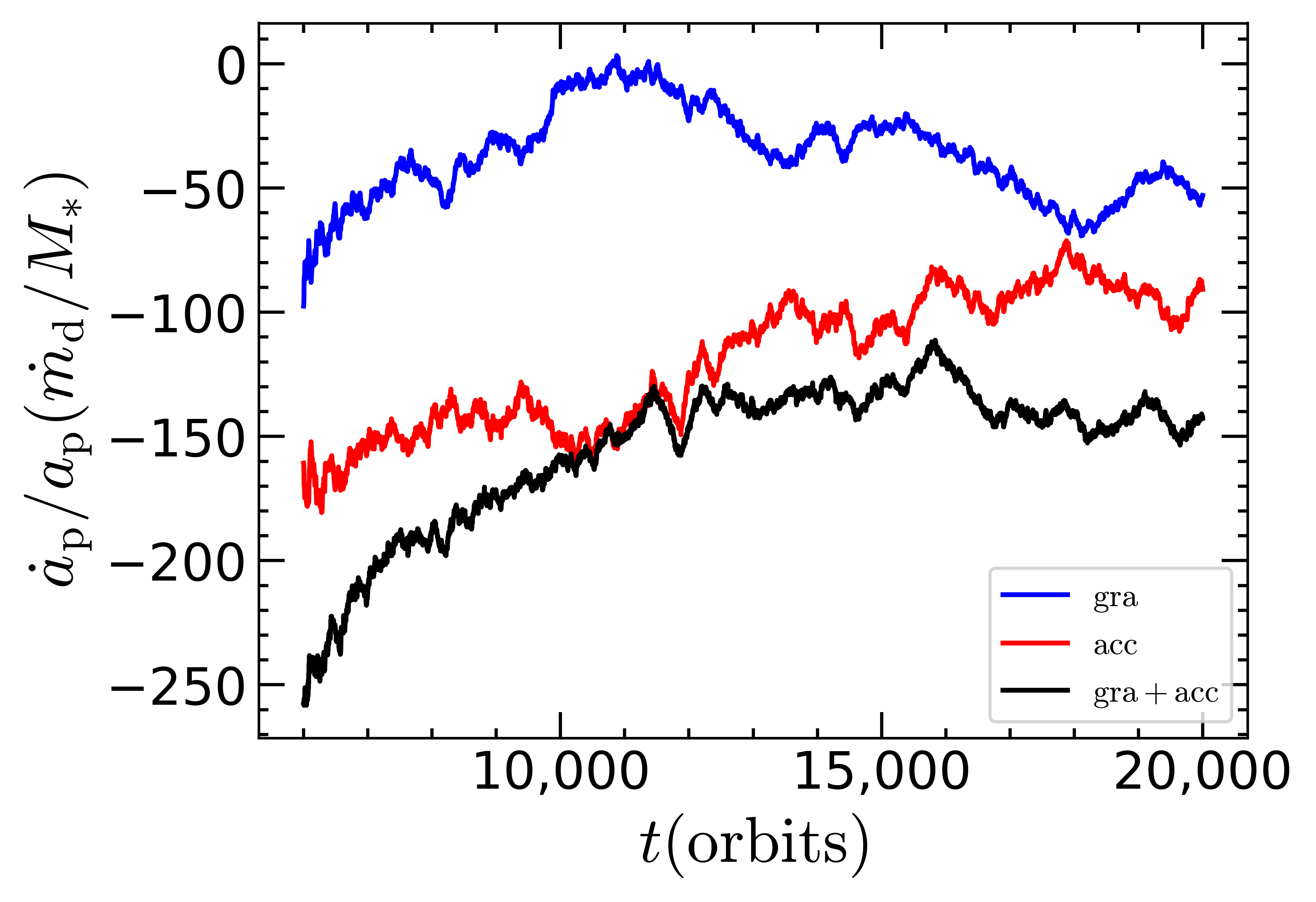}
\end{adjustwidth}
\caption{The semi-major axis evolution due to the gravitational (blue lines), accretional (red lines) and total (black lines) torques in isothermal (left panel) and $\beta=100$ (right panel) cases. For the case of $\beta=100$, a running-time averaged over 500 orbits is carried out to smooth out the short timescale variability. 
}
\label{fig:adot_time}
\end{figure}

\subsection{The migration of the planet} \label{subsec:re_mig}

\begin{figure}
\centering
\includegraphics[width=0.8\textwidth]{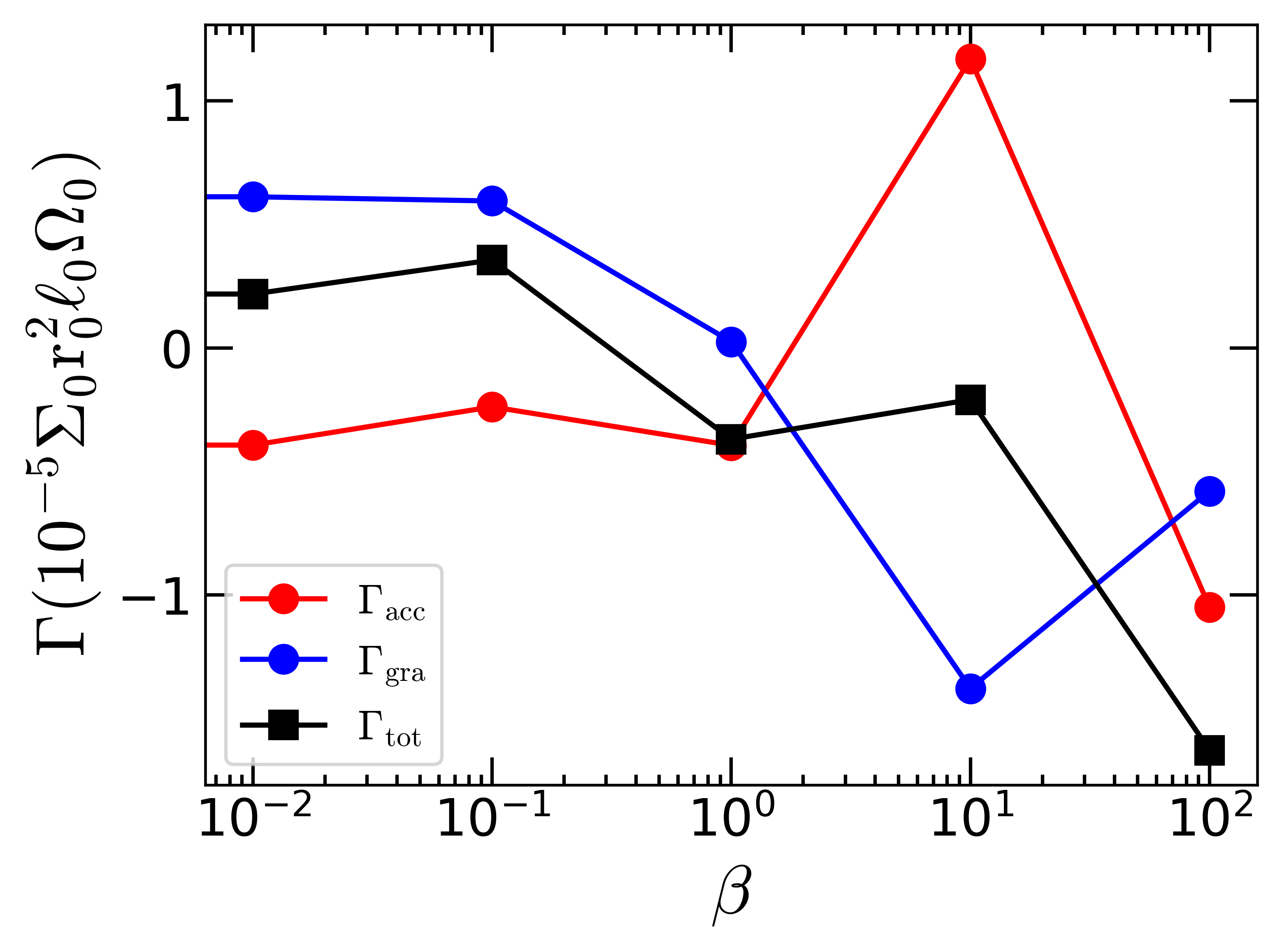}
\caption{The torque exerted on the planet under different cooling timescales in code unit, where $\ell_{0}=r_{0}^{2}\Omega_{0}$ is the specific angular momentum of the disk at $r_{0}$. The blue and red circle points represent the gravitational torque and accretional torque respectively. The black square points represent the sum of the two torques. The positive (negative) torque means that it drives the planet to migrate outward (inward). 
\label{fig:torque_beta}}
\end{figure}

To study the effects of thermodynamics on the migration of the planets, we calculate the torques exerted onto the planet. The torques can be divided into two parts, the first part is from the gravitational force from the PPD, the other one is from the mass and angular momentum transfer in the sink hole area, even though we do not actively add them onto the planet in our simulations \citep{Li2021a,Li2021b,LiR2022}. Considering the small mass ratio between the planet and star ($q=10^{-3}$), the gravitational torque and accretional torque associated with migration can be calculated as follows \citep{Li2024},

\begin{equation} \label{equ:t_gra}
\Gamma_{\rm gra} = \vec{r_{\rm p}} \times \int \Sigma \nabla \Phi_{\rm p} {\rm d}S,
\end{equation}
and

\begin{equation} \label{equ:t_acc}
\Gamma_{\rm acc} = \vec{r_{\rm p}} \times \int_{\delta r < r_{\rm acc}} (\vec{v}-\vec{v}_{\rm p}) {\rm d}\dot{m}_{\rm p},
\end{equation}
where $\Phi_{\rm p}$ is the gravitational potential of the planet, $\vec{v}$ is the velocity of the gas within the sink hole radius in the inertial frame. The migration rate for the planet's semi-major axis $a_{\rm p}$ can be expressed as

\begin{equation} \label{equ:adot}
\frac{\dot{a}_{\rm p}}{a_{\rm p}}=\frac{2\ell_{\rm p}}{\ell_{\rm p}}=\frac{2(\Gamma_{\rm gra}+\Gamma_{\rm acc})}{m_{\rm p}\ell_{\rm p}},
\end{equation}
where $\ell_{\rm p}=\sqrt{GM_{*}a_{\rm p}}$ is the specific angular momentum of the planet. Here we have neglected the accretion term onto the central star $-\dot{M}_{*}/M_{*}$ as this is usually negligible.

The migration rates of the semi-major axis $a_{\rm p}$ for two typical models are shown in Figure~\ref{fig:adot_time}. It is evident that both the isothermal case and the one with $\beta=100$ achieve a quasi-steady state at approximately 20,000 orbits. In the isothermal case, the planet migrates outward, whereas it migrates inward in the case of $\beta=100$.

We present the results for the migration torques for all different models in Figure \ref{fig:torque_beta}. The migration torque is usually dominated by the gravitational component except for the cases with $\beta\gtrsim 10$, where the accretional component becomes comparable to the gravitational torque; however, it is still not the primary factor determining the direction of migration.
It can be seen that as  the cooling timescale increases, the gravitational torque $\Gamma_{\rm gra}$ and also the total torque $\Gamma_{\rm tot}$ drop monotonically from positive to negative, with $\Gamma_{\rm tot}>0$ for $\beta\lesssim 1$ and $\Gamma_{\rm tot}<0$ for $\beta1$. This suggests that when the cooling timescale exceeds the local orbital timescale (i.e., $\beta \gtrsim 1$), migration transitions to inward. The positive torque for an accreting Jupiter-mass planet in the isothermal case with a high disk viscosity is consistent with the results from \citet{Li2024,Laune2024}.

In the type I/II regime for non-accreting planets, the inward migration rate is $\dot{a}_{\rm p}/a_{\rm p} \simeq
-4 q/h_{0}^{2}\Sigma_{\rm p}r_{\rm 0}^{2}\Omega_{0}/M_{*}$ $\simeq-500\dot{m}_{\rm d}/M_{*}$ \citep{Tanaka2002}, where the $\Sigma_{\rm p}$ is the surface density at the planetary orbit after considering the gap opening effect \citep{Kanagawa2018}. We can see that the inward migration rate in the high $\beta$-cooling cases is a factor of $\sim3$ slower than that of the non-accreting planet.
In the adiabatic limit for the non-accreting planet, the migration torque can be positive for our density and temperature profiles due to  the non-linear corotation torque \citep{Paardekooper2010}. 
With a long $\beta$-cooling timescale, which is close to the adiabatic case, the negative gravitational torque on the accreting planet is due to the suppression of the positive corotation torque associated with accretion.

\begin{figure*}
\begin{adjustwidth}{-\extralength}{0cm}
\centering
\includegraphics[width=0.9\textwidth]{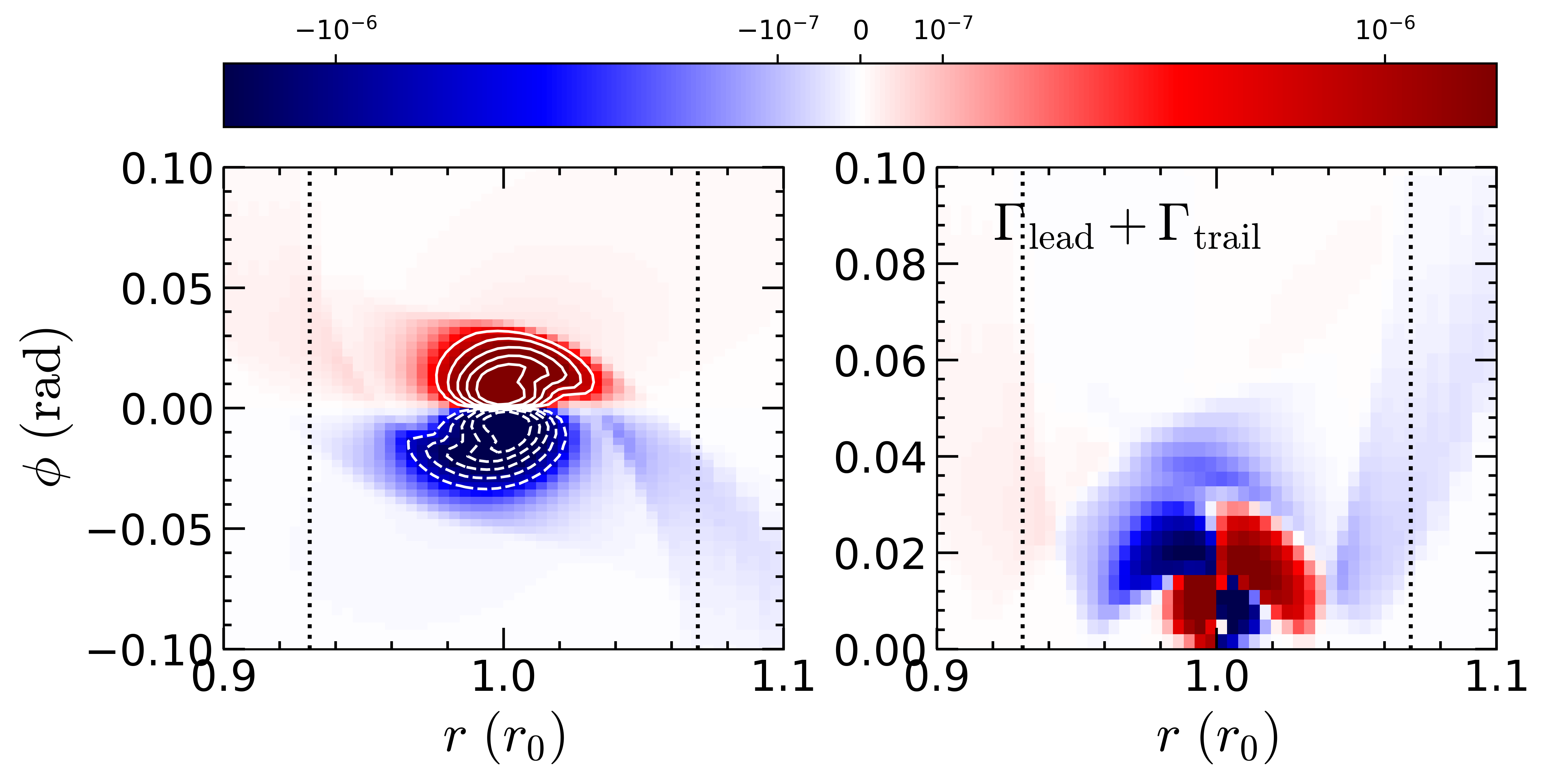}
\includegraphics[width=0.9\textwidth]{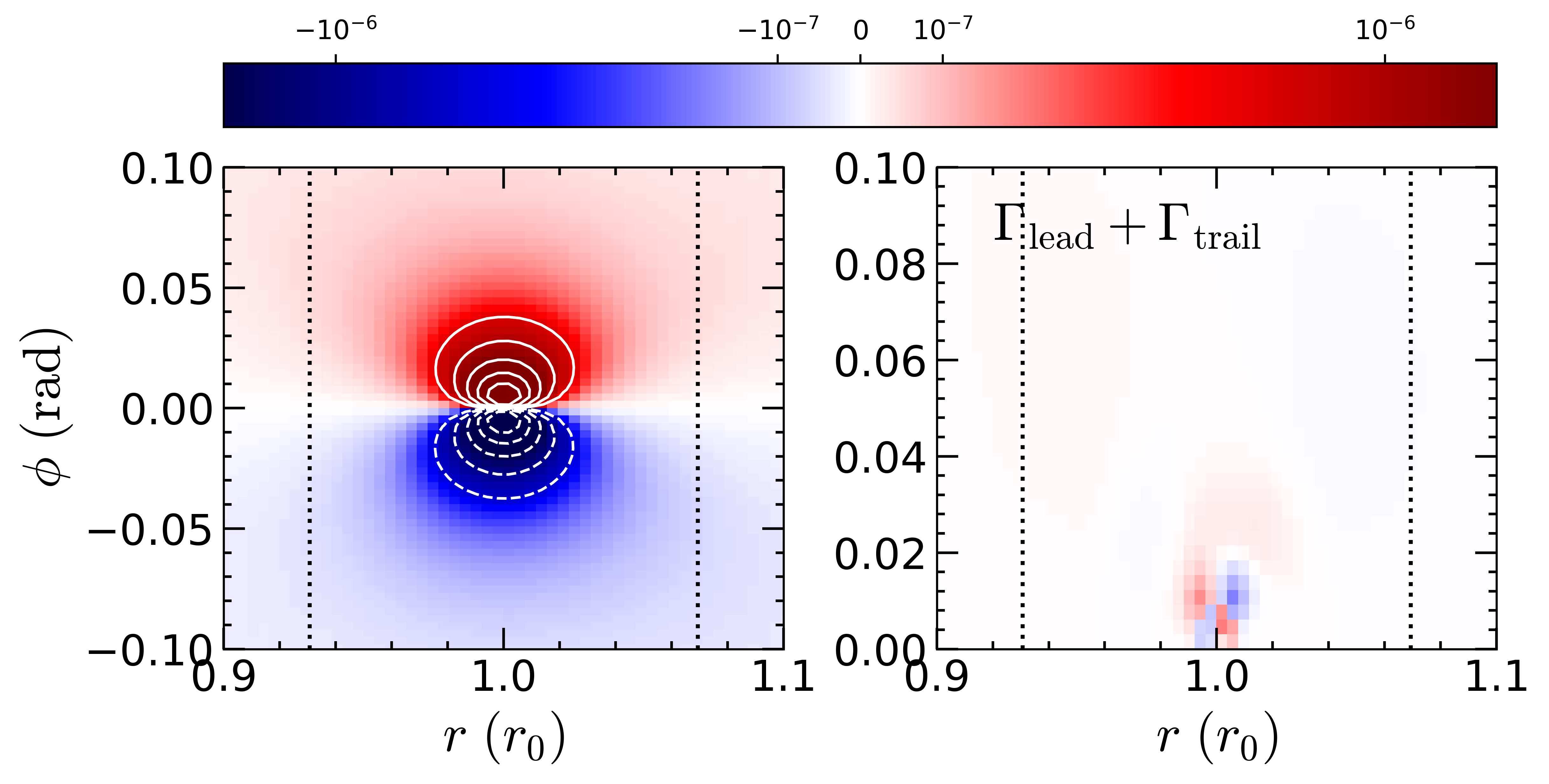}
\end{adjustwidth}
\caption{Spacial distribution of gravitational torques in the vicinity of the planet around 20000 orbits. The upper panel represents the isothermal condition and the lower panel shows the $\beta=100$ case. A time-averaging over 500 orbits is done for the $\beta=100$ case. The contour in the left panels represented by solid and dashed lines being the same magnitude of the torques but with different signs. The vertical dotted lines correspond to the region of $1\ r_{\rm h}$ from the planet. The right panels show the net torque by summing the leading (upper) horseshoe and trailing (lower) horseshoe region, i.e., $\Gamma(r,\phi)+\Gamma(r, -\phi)$.}
\label{fig:torque2d}
\end{figure*}

\subsection{How Thermodynamics Affects Migration} \label{subsec:dis_mig}

The gravitational torque could be the dominant factor that determines the planet migration. To understand what causes the transition of the migration, we explore the influence of thermodynamics on the gravitational torque. As the regions where the Lindblad torque dominates and the corotation torque dominates overlap with each other, we do not distinguish their contributions in this paper. Following \citet{Li2024}, we plot the spatial torque distribution in the vicinity of the planet and the results are shown in Figure~\ref{fig:torque2d}. The CPD structure is clear in the isothermal case in which the leading horseshoe region  provides positive torque (red color region) while the trailing region provides negative torque (blue color). In the $\beta=100$ case, the disk-like structures break down due to thermal pressure and the torque within the Hill sphere is highly symmetrical with respect to the $\phi=0$ plane. We then calculate the asymmetrical part of the torque defined as $\Gamma_{\rm asy}=\Gamma(r,\phi)+\Gamma(r,-\phi)$ as shown in the right panels in the figure. It can be seen that in the CPD region for the isothermal case, the positive torque from the $\phi>0$ hemisphere is stronger than the negative torque from the $\phi<0$ hemisphere so the net torque is positive, as expected from the asymmetric spiral arms feeding from the global disk into the Hill sphere \citep{Li2024}. While for the high $\beta$-cooling case, the net positive torque in the vicinity of the planet almost vanishes, due to that the asymmetric disk-like structure is replaced by a spherical-like envelope within the Hill sphere.

The asymmetry of the gravitational torque comes from the asymmetry of the density wave structure in the vicinity of the planet. We plot the surface density distribution around the planet in Figure~\ref{fig:CPD_2d}. The asymmetry in the spiral arms feeding the planet occurs in the isothermal case, but vanishes in the $\beta=100$ case. In the accretion process of the planet, the materials from the outer regions of the disk approach the planet through outer horseshoe tracks and accumulate at the CPD region. Part of the materials are accreted onto the planet with the left materials moving to the inner horseshoe tracks and will enter the CPD region again after one orbit. Therefore, the asymmetries in density and torque originate from the accretion process of the planet. In contrast, for the $\beta=100$ case, the CPD structure is destroyed and supported by the thermal pressure gradient, which tends to be nearly isotropic, such that the accretion tends to be Bondi-like. 

\begin{figure}
\centering
\includegraphics[width=0.8\textwidth]{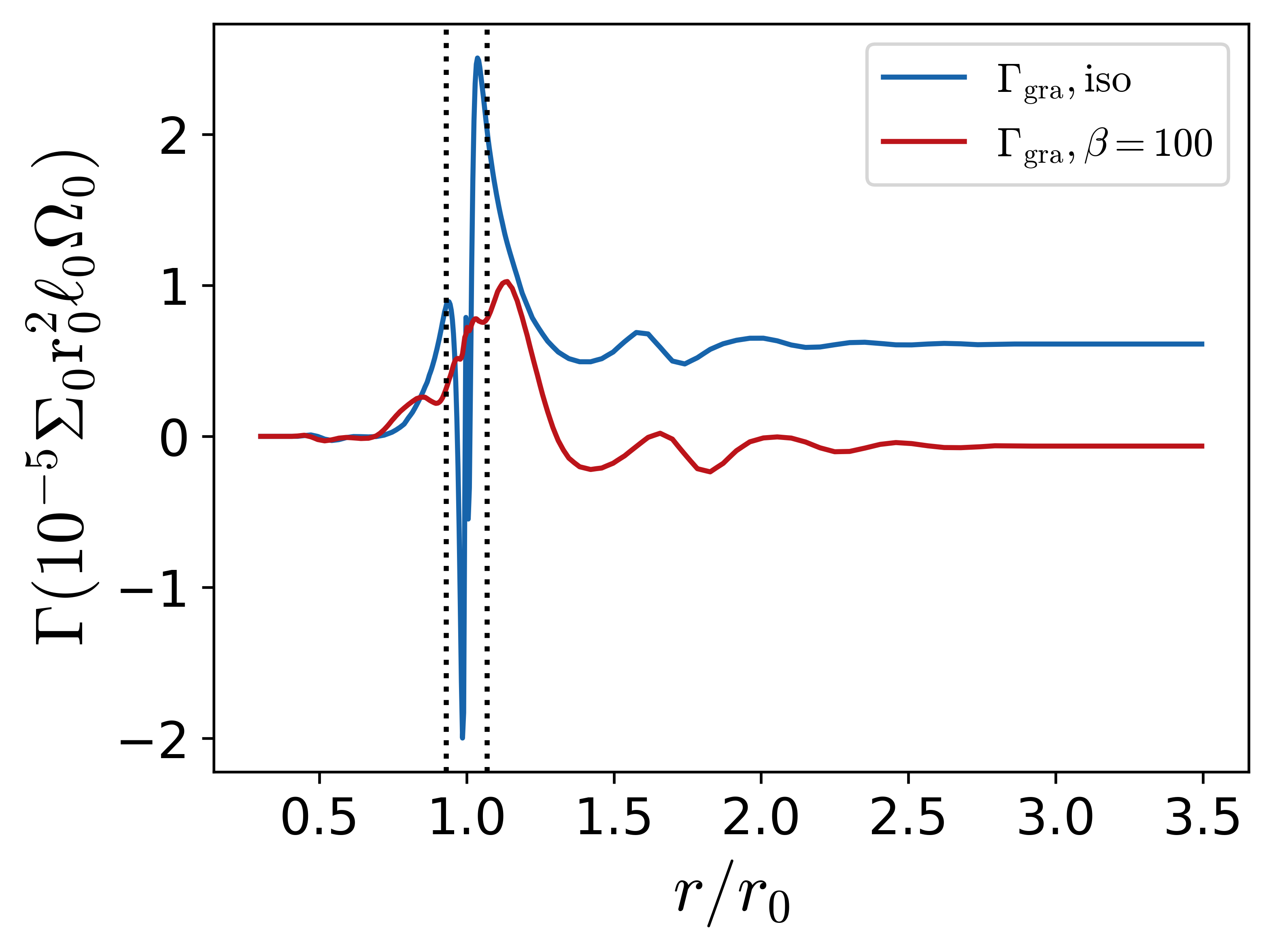}
\caption{The cumulative radial distribution (integration from $r_{\rm in}$ to $r$) of the gravitational torque exerted onto the planet, where the blue (red) line represents the  isothermal ($\beta=100$) case. The torque for $\beta=100$ is time-averaged over 500 orbits around $20000$ orbits. The two vertical dotted lines show the Hill radius. The torque is in code unit with $\ell_{0}=r_{0}^{2}\Omega_{0}$ being the specific angular momentum of the disk at $r_{0}$.
\label{fig:torque_r}}
\end{figure}

The cumulative radial torque profiles by azimuthally averaging the 2D spatial gravitational torque distribution $\int_{>r}\int_{\phi}\Gamma(r,\phi){\rm d} r{\rm d} \phi$ as a function of radius for two different models are shown in Figure~\ref{fig:torque_r}. We can see that the torque outside the Hill sphere ($r<r_{\rm p}-r_{\rm h}$, $r>r_{\rm p}+r_{\rm h}$), which is dominated by the Lindblad resonance, is negative both in isothermal and long cooling timescales. These differential Lindblad torque usually drives the planet migrate inward. 
However, as shown in \citet{Li2024}, the gravitational torque within the planet's Hill sphere is significantly positive in cases close to the isothermal EOS, allowing it to counterbalance the negative differential Lindblad torque. This is found to be associated with the asymmetric structures in the CPD region as discussed above.
As the cooling timescale becomes longer (e.g., $\beta=100$), a rotationally supported CPD disappears, the positive gravitational torque from the CPD region thus becomes too weak to balance the negative differential Lindblad torque, in line with Figure~\ref{fig:torque2d}. As a result, the planet migrates inward.

\section{Conclusions and Discussion} \label{sec:conclusions}

In this work, we perform a series of 2D global hydrodynamical simulations using \texttt{Athena++} to study the impact of thermodynamics on gas giants' concurrent accretion and migration embedded within their natal disks. 
The disk is evolved over the viscous timescale with a desired mass flux from the outer boundary, allowing the system to reach a quasi-steady state in a self-consistent manner.
A parameterized thermal relaxation process with $\beta$-cooling is implemented to simulated the critical role of disk thermodynamic on the accretion and migration dynamics of the embedded object.
The planetary accretion is modeled as a sink particle, with a high resolution around the planet to resolve the CPD region. Our findings are summarized as follows:

\begin{itemize}
    \item The planetary accretion rates for varying cooling timescales are primarily constrained by the disk supply rate from the outer disk, exhibiting a slight decrease with increasing cooling timescales, even though different thermodynamical states have a significant impact on the CPD structures.
    \item The CPD is much smaller in size and less rotationally supported with longer cooling timescales. When the cooling timescale is an order of magnitude longer than the local dynamical timescale (e.g., $\beta\gtrsim10$), the rotation of the CPD can even shift to a retrograde orbit, which could strongly influence satellite formation within the CPD.
    \item The planetary accretion is chiefly driven by spiral shock dissipation in the CPD region. We observe that the Reynolds stress coefficient varies significantly with different cooling timescales, following a trend similar to that of the effective accretion coefficients, making it a dominant factor in controlling the accretion process onto the planet.
    \item We confirm that in an isothermal disk, the planet migrates outward for a high disk viscosity, consistent with recent findings by \citet{Li2024,Laune2024}. However, there is a tendency for inward migration as the cooling timescale increases ($\beta \geq 1$), different from the general trend where non-accreting planets migrate outward in the adiabatic limit. 
    \item The transition in migration under the longer cooling timescales is closely linked to the gravitational torque from the co-orbital region of the planet. In longer cooling timescales, the positive gravitational torque from this region is significantly suppressed due to the absence of a rotationally supported CPD. These results highlight the critical role of thermodynamics in the accretion and migration of planets in PPDs.    
\end{itemize}

The physical process in this work could be also relevant to the dynamics of the accreting stellar-mass black holes/compact objects embedded in AGN disks. Their dynamical evolution in AGN disks could provide be an important channel for binary black hole mergers detected by LIGO/Virgo/KAGRA gravitational wave observations \citep[e.g.,][]{Li2021b,Li2022b,LiR2022}.
In this case, the mechanical and radiative feedback from the embedded compact objects could be significant to modify the circumsingle disk structures, and thus their dynamical evolution. 
In addition, the inner solar system are subjected to extreme radiation environments from young stars \citep{Ribas2005}. Massive asymptotic giant branch stars could also play a significant role in delivering short lived radionuclides to heat rock-forming materials in the disk \citep{TrigoRodriguez2009}.
All of these would necessitate a more self-consistent approach to cooling and heating with radiative hydrodynamics simulations for the CPD, rather than relying on the $\beta$-cooling prescription.
Second, a laminar viscous flow with a $\alpha-$viscosity is assumed in our work. Magnetic turbulence in disks, which could modify the coherent spiral waves lunched by the embedded objects, may also play an important role in the dynamical evolution of embedded objects. Third, our focus has been primarily on a parameter space involving a Jupiter mass planet embedded in a highly viscous disk fixed in a circular orbit. A systematic exploration of different planet masses, disk viscosities, disk scale heights, and planetary eccentricities is still needed, which could be crucial for future population synthesis studies. Lastly, our simulations are limited to 2D cases. While our previous studies have shown that 3D simulations yield similar accretion and migration dynamics as 2D cases with an isothermal EOS \citep{Li2024}, the extent of the 3D effects for our adiabatic equation of state with varying $\beta$-cooling remains unclear. All these aspects will be addressed in future work.

\vspace{6pt}

\authorcontributions{
Conceptualization, methodology, supervision, funding acquisition, Y.P.L.; writing---original draft preparation, H.W.; software, validation, formal analysis, visualization, investigation, resources, writing---review and editing, H.W., Y.P.L.; 
All authors have read and agreed to the published version of the manuscript.
}

\funding{This work is supported in part by the Natural Science Foundation of China (grants Nos. 12373070, and 12192223), the Natural Science Foundation of Shanghai (grant No. 23ZR1473700).}

\dataavailability{All the data that are used in this article are available from the corresponding author upon reasonable request.} 




\acknowledgments{We thank the referees for many constructive suggestions. We also thank beneficial discussions with Zhaohuan Zhu. The calculations have made use of the High Performance Computing
Resource in the Core Facility for Advanced Research Computing
at Shanghai Astronomical Observatory. }

\conflictsofinterest{The authors declare no conflicts of interest.} 



%

\appendixtitles{no} 
\appendixstart
\appendix
\section[\appendixname~\thesection]{} \label{subsec:con_test}

Here we test the convergence of the accretion prescription and numerical resolution for a model in the transition region (i.e., $\beta=1$),  mainly focused on their impact on the accretion dynamics.
There are two parameters for the planetary accretion,  the removal rate $f$, and the accretion radius (or sink hole radius) $r_{\rm acc}$.

\begin{figure}
\centering
\includegraphics[width=0.8\textwidth]{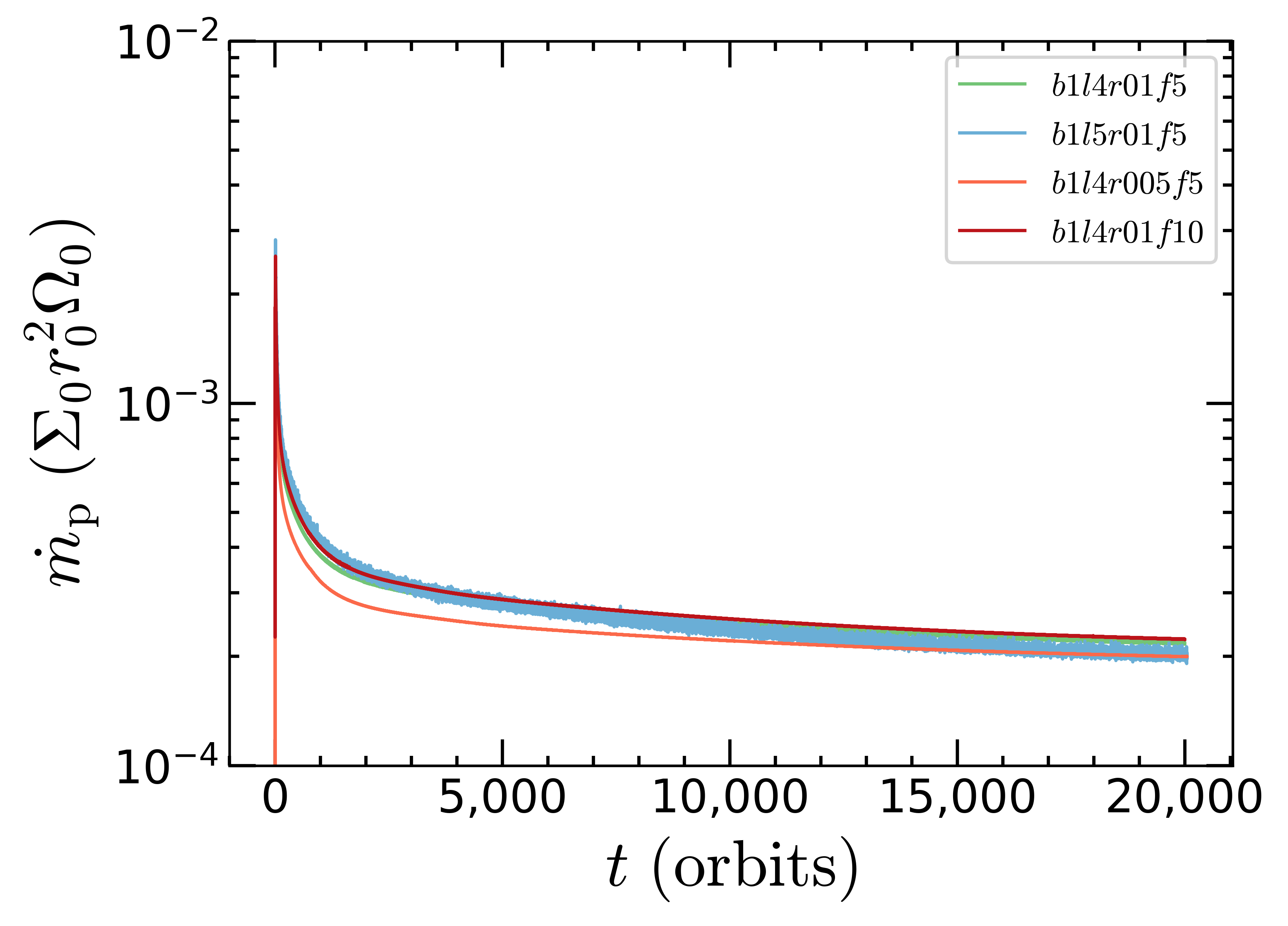}
\caption{The accretion rates for different parameters. The number following the letter 'b' indicates the value of $\beta$, 'l' refers to the mesh refinement level, 'r' denotes the sink hole radius, which is the same as the softening radius, and 'f' represents the removal rate of the sink hole region, e.g., 'b1l4r01f5' indicates $\beta=1$, 4 levels of mesh refinement, $r_{\rm acc}=0.1\ r_{\rm h}$, and $f=5$. It is clear that all of these parameters have a negligible effect on planetary accretion rates.}
\label{fig:con_test}
\end{figure}

The sink hole radius $r_{\rm acc}$ should be small enough to represent the size of the accreting object accurately. However, using a sink hole radius comparable to that of a Jupiter-mass planet necessitates an extremely high resolution around the planet in our global simulations, which is impractical even with mesh refinement. Therefore, it is advisable to select an appropriate value for $r_{\rm acc}$ that does not significantly affect the dynamical accretion of the embedded object. By comparing the simulation results with $r_{\rm acc}=0.1r_{\rm h}$, $0.05r_{\rm h}$, and different removal rates $f$, we observe no significant differences in the accretion rate, as illustrated in Figure~\ref{fig:con_test}. This indicates good convergence with the accretion prescription.

In addition, we have conducted convergence tests at various numerical resolutions in the vicinity of the planet. With 4 and 5 levels of mesh refinement around the planet, we found that the differences in accretion rates among these cases are negligible. 
Similar tests were performed for different accretion prescriptions and global numerical resolutions (up to $n_{r}=1800$, and $n_{\phi}=2500$) without any mesh refinement using \texttt{FARGO3D} \citep{Benitez-Llambay2016}. All of these indicate satisfactory convergence for our simulation results.

\begin{adjustwidth}{-\extralength}{0cm}

\reftitle{References}


\bibliographystyle{plain}



%


\PublishersNote{}
\end{adjustwidth}

\end{document}